# Seyferts on the edge: polar scattering and orientation-dependent polarization in Seyfert 1 nuclei

J. E. Smith[1*], A. Robinson[1†], D. M. Alexander[2], S. Young[1], D. J. Axon[1†], Elizabeth A. Corbett[3]

*1 Department of Physical Sciences, University of Hertfordshire, Hatfield, Hertfordshire, AL10 9AB, UK*
*2 Institute of Astronomy, Madingley Road, Cambridge, CB3 0HA, UK*
*3 Anglo-Australian Observatory, PO Box 296, Epping, NSW 1710, Australia*

## ABSTRACT

We have identified 12 Seyfert 1 galaxies that exhibit optical polarization spectra similar to those of Seyfert 2 galaxies in which polarized broad-lines are detected. We present new spectropolarimetric observations of 3 of them: Was 45, Mrk 231 and NGC 3227. These objects appear to be polarized as a result of far-field scattering in the polar illumination cones of the circum-nuclear torus. We estimate that they represent between 10 and 30 per cent of the Seyfert 1 population; they are found amongst all the main spectroscopic sub-types, including narrow-line Seyfert 1 galaxies. We have shown elsewhere that Seyfert 1 nuclei more commonly have polarization characteristics that can be attributed to scattering by a compact 'equatorial' scattering region located inside the torus. We propose that both equatorial and polar scattering regions are present in all Seyfert galaxies and argue that the observed range of polarization properties can be broadly understood as an orientation effect. In this scheme, polar-scattered Seyfert 1 galaxies represent the transition between unobscured (the majority of type 1) and obscured (type 2) Seyferts. They are viewed through the upper layers of the torus and are thus subject to moderate extinction ($A_V \approx 1$–4 mag) sufficient to suppress polarized light from the equatorial scattering region, but not the broad wings of the Balmer lines. The orientation of the polarization position angle relative to the radio source is broadly consistent with the two-component scattering model. More generally, we find that amongst Seyferts 1 galaxies, parallel, perpendicular and intermediate orientations of the polarization PA relative to the radio axis occur roughly in the proportions 2:1:1.

## 1  INTRODUCTION

The detection of polarized broad-lines in the nucleus of the Seyfert 2 galaxy NGC 1068 (Antonucci & Miller 1985) was a key step in the development of the current unification scheme for Seyfert galaxies. This holds that Seyfert 1 and 2 galaxies are intrinsically the same type of object, but viewed from different orientations. It is postulated that the central continuum source and broad emission line region (BLR) are surrounded by an optically and geometrically thick torus of molecular gas and dust, the orientation of which relative to the line-of-sight determines the observed spectrum. In Seyfert 2 galaxies, the direct view of the active nucleus is blocked by the torus. However, radiation escapes from the nucleus in conical beams aligned with the poles of the torus. Some of this radiation is scattered towards us and as a result, Seyfert 2 galaxies may exhibit the spectroscopic properties of Seyfert 1 galaxies in polarized light.

* E-mail: jsmith@star.herts.ac.uk
† Current address: Department of Physics, Rochester Institute of Technology, 85 Lomb Memorial Drive, Rochester, NY 14623-5603, USA

Polarized broad-lines have indeed been found in many other Seyfert 2 galaxies in addition to NGC 1068 (e.g. Miller & Goodrich 1990; Tran, Miller & Kay 1992; Young et al. 1996; Heisler, Lumsden & Bailey 1997; Lumsden et al. 2001; Tran 2001). In NGC 1068 itself, imaging polarimetry has revealed a bi-polar scattering region with a centro-symmetric polarization pattern, consistent with illumination by conical beams from a hidden central source (e.g. Capetti et al. 1995a; Packham et al. 1997; Simpson et al. 2002). More generally, in Seyfert 2 galaxies, the optical polarization position angle (PA) is almost always perpendicular to the projected radio source axis (e.g. Antonucci 1983; Brindle et al. 1990). This is precisely what is expected for scattering of light in the 'illumination cone' of the torus if the radio source is co-aligned with the torus axis.

Therefore, the basic polar scattering geometry postulated by the unification scheme seems to be generally valid for Seyfert 2 galaxies. In contrast, the optical polarization properties of most Seyfert 1 galaxies are *not* consistent with polar scattering. Several previous studies have found that the optical polarization PA is more often *aligned* with the radio axis in these objects (e.g. Antonucci 1983, 2001; Martel 1996, hereafter M96; Smith et al. 2002, hereafter S02). At least some of the scattered light emerging from the nucleus must therefore follow a different path to that in the Seyfert 2 galaxies, suggesting, in turn, that the simplest unification geometry including only polar scattering, is incomplete. Evidently, we require an additional source of polarized light in Seyfert 1 galaxies, which has a different scattering geometry and moreover, is not seen in Seyfert 2 galaxies.

In our recent detailed study of the optical polarization spectra of Seyfert 1 nuclei (Smith 2002; S02), we argued that a number of commonly exhibited polarization characteristics are best explained by a model in which broad Balmer-line emission originates in a rotating disc and is scattered by material in the equatorial plane of the disc (see also Goodrich & Miller 1994, hereafter GM94; Cohen et al. 1999; Cohen & Martel 2001). This equatorial scattering model can simultaneously account for both the position angle rotations and the percentage polarization structures that are observed across the broad H$\alpha$ lines in many objects (Smith 2002; Smith et al. 2004, in preparation, hereafter S04). The model predicts that the average polarization PA of the broad H$\alpha$ line, and that of the continuum, is aligned with the rotation axis of the line-emitting disc and therefore, the axes of the torus and radio source. We can thus account for the alignment between the optical polarization PA and the radio source, which is observed in the majority of objects.

However, to further complicate the picture, several Seyfert 1 galaxies show optical polarization *orthogonal* to their radio source axes (Antonucci 2001; see also S02). This is most readily explained by polar scattering, as for Seyfert 2 galaxies. Furthermore, we also identified in S02 a small number of Seyfert 1 galaxies whose polarization spectra are very similar to those of Seyfert 2 galaxies in which polarized broad-lines are detected (hereafter referred to as PBL Seyfert 2 galaxies). Whilst equatorial scattering seems to explain the optical polarization in many Seyfert 1 galaxies, this is strong evidence that polar scattering not only occurs in Seyfert 1 galaxies (as would be expected in the unification scheme) but actually dominates over equatorial scattering in some cases. These objects are of particular interest in the context of the unification scheme, since they appear to have scattering geometries similar to Seyfert 2 galaxies, despite exhibiting Seyfert 1 spectra in total light

In this work we investigate in more detail the nature of those Seyfert 1 galaxies that share the polarization characteristics of PBL Seyfert 2 galaxies. We have identified 3 such objects in our original sample (S02) and report here the discovery of a fourth, NGC 3227. Several more candidates have been identified from the literature. We also present new observations of two previously studied objects, Mrk 231 (GM94; M96) and Was 45 (one of the sample discussed in S02). We hereafter refer to these objects collectively as 'polar-scattered' Seyfert 1 galaxies.

In Section 2 we discuss the polarization properties of the individual objects and how they differ from the majority of Seyfert 1 galaxies. In Section 3, we summarise the mechanisms that are believed to produce the optical polarization spectra of PBL Seyfert 2 galaxies. In Section 4 we outline a two-component scattering model including both equatorial and polar scattering regions. Using a scattering code, we show that the inclination of the symmetry axis to the line-of-sight cannot, by itself, govern which component dominates the observed polarization. We argue instead, that polar-scattered Seyfert 1 galaxies are objects in which the line-of-sight to the nucleus passes through the upper layers of the torus and estimate the amount of extinction needed for the polar scattering region to dominate the observed polarization. Finally, in Section 5, we argue that the range in polarization properties observed amongst Seyfert galaxies can be broadly understood in terms of an orientation sequence consistent with the unification scheme. We also discuss the prevalence of polar-scattered objects among the Seyfert 1 population and orientation of the polarization PA relative to the radio source for a sample of 22 objects in which the axis of the latter can be determined.

## 2 POLARIZATION PROPERTIES OF POLAR-SCATTERED SEYFERT 1 GALAXIES

We have recently studied the optical polarization properties of a sample of 36 Seyfert 1 galaxies (S02). Of these, 20 exhibit variations in polarization with wavelength across the broad H$\alpha$ line (either a rotation in position angle, $\theta$, and/or changes in degree of polarization, $p$), which indicate that the Seyfert nucleus is intrinsically polarized. The remainder are either weakly polarized ($p \leq 0.3$ per cent), or may be significantly affected by foreground interstellar polarization.

In the 20 intrinsically polarized Seyfert 1 galaxies, the polarization structure across the broad H$\alpha$ line differs widely from object to object. However, it is possible to discern certain characteristics that are present to differing degrees in most objects. These are (i) a swing in position angle across the broad H$\alpha$ profile and (ii) a dip in polarization in the core of the profile, flanked by polarization peaks in the wings. In S02, we argued that both of these characteristics are naturally explained if the H$\alpha$ emission comes from a rotating disc and is scattered in a closely

**Table 1.** Log of observations. Exposure times are the sum of equal exposures at each of the 4 waveplate positions.

| Object | z | Telescope | Date | Exp. (s) |
|---|---|---|---|---|
| Mrk 231 | 0.042 | WHT | 21/2/91 | 2×2000 |
| NGC 3227 | 0.004 | WHT | 27/4/02 | 2000 |
| Was 45 | 0.025 | WHT | 26/4/02 | 4000 |

surrounding co-planar ring, both structures residing in the equatorial plane of the torus.

However, there are other objects in the sample whose *percentage polarization* spectra are quite distinct from those showing characteristics attributable to equatorial scattering. The polarization spectra of these objects are, in fact, much more similar to those of PBL Seyfert 2 galaxies. The prototype is Fairall 51, whose polarization spectrum has also recently been discussed by Schmid et al. (2001). We have identified two other candidates in our original Seyfert 1 sample, NGC 4593 and Was 45, and report here the discovery of a third, NGC 3227. Another object with similar polarization properties, ESO 323–G077, has recently been discussed by Schmid, Appenzeller & Burch (2003). A literature search has revealed several more good candidates: Mrk 231, Mrk 704 and Mrk 376 (GM94), Mrk 1218 (Goodrich 1989a) Mrk 766, *IRAS* 1509–2107 and Mrk 1239 (Goodrich 1989b). We also present previously unpublished spectropolarimetry of Mrk 231 and new observations, covering a wider spectral range, of Was 45.

### 2.1 Observations

Spectropolarimetric observations of Mrk 231 were obtained using the red arm of the ISIS dual beam spectrograph at the 4.2-m William Herschel Telescope (WHT) on the night of 1991 February 21. The spectrograph was used in standard polarimetry mode, with the halfwave plate inserted into the beam above the slit unit, and a calcite slab analyser below. A comb dekker was used to prevent overlap between the orthogonally polarized spectra produced by the analyser. The total exposure time was 4000 seconds. The detector was a 1242×1152 pixel EEV CCD, which, with the R158R grating (158 lines mm$^{-1}$) and a 1 arcsec slit, gave an effective spectral resolution ≈7Å.

**Table 2.** Measured average polarization and estimated line-of-sight polarization for new observations.

| Object | Broad H$\alpha$ range (Å) | $p$ (per cent) | | $\theta$ (deg) | | Gal. $E(B-V)$ (mag) | Interstellar $p$ (per cent) | |
|---|---|---|---|---|---|---|---|---|
| | | Cont. | H$\alpha$ | Cont. | H$\alpha$ | | Typical | Max |
| Mrk 231 | 6660–6940 | 2.95±0.04 | 3.40±0.09 | 95.0±0.4 | 92.7±0.7 | 0.010 | 0.03 | 0.09 |
| NGC 3227 | 6520–6660 | 0.66±0.02 | 1.02±0.03 | 122.4±1.0 | 121.3±0.7 | 0.023 | 0.07 | 0.21 |
| Was 45 | 6660–6830 | 0.57±0.03 | 2.46±0.07 | 149.4±1.7 | 142.2±0.8 | 0.018 | 0.05 | 0.16 |

The new observations of Was 45 and NGC 3227 were also made at the WHT, on the nights of 2003 April 26 and 27, respectively. The red arm of the ISIS spectrograph was used with the polarimetry optics, a 2148×4700 pixel Marconi CCD detector and the R316R (316 lines mm$^{-1}$) grating. This set-up gave an unvignetted spectral range of 2300Å and a dispersion of 0.84 Å pixel$^{-1}$. A slit width of 1 arcsec was chosen to maximise throughput (the seeing was in the range 0.6–1.0 arcsec) and since this projects to 4 pixels at the detector the effective spectral resolution was ≈3.4 Å. These spectra therefore have resolution comparable to the WHT data previously presented in S02, but much greater wavelength coverage, allowing us to include both the Hα and Hβ emission lines in the spectrum.

The target sources and standard stars were observed using the standard spectropolarimetry procedure of taking equal exposures at half-wave plate angles of 0, 22.5, 45 and 67°. Polarized standard stars were observed to determine the zero point of the polarization PA. Unpolarized standards were observed to check the instrumental polarization. Spectrophotometric standards were also observed to allow flux calibration of the spectra. A summary log of the observations is given in Table 1.

The data were reduced and analysed using the methods described in S02. We note that *p* is subject to a positive bias (e.g. Serkowski 1958; Clarke & Stewart 1986) which, when the signal-to-noise ratio is low, may lead to spurious features in the polarized flux spectrum formed by multiplying the total flux ($F_\lambda$) by $p(\lambda)$. This representation of the polarized flux also requires careful interpretation when large polarization PA rotations are present. To avoid such problems, it is sometimes preferable to form the polarized flux spectrum using the rotated Stokes Parameters (e.g. see Tran 1995; Hines et al. 1999a). However, the data presented here are of relatively highly polarized objects and have a good signal-to-noise ratio. Furthermore, one of the main distinguishing characteristics of the Seyfert 1 galaxies discussed in this work is that the polarization PA is approximately constant with wavelength. In these circumstances ($F_\lambda$) × $p(\lambda)$ is an entirely adequate representation of the polarized flux spectrum and hence, for consistency with S02, we use this form here.

The polarization spectra ($F_\lambda$, percentage polarization $p(\lambda)$, polarization position angle $\theta(\lambda)$ and polarized flux $p(\lambda) \times F_\lambda$) of Fairall 51 and NGC 4593, previously presented in S02, are reproduced in Figs. 1 and 2, for convenience. The new data on Was 45, Mrk 231 and NGC 3227 are presented in Figs. 3–5. Wavelength-averaged values of *p* and *θ* for both the broad Hα line and adjacent continuum windows were measured from the new spectra and are listed Table 2. These measurements were made using the procedures outlined in S02; the emission-line polarization being determined following continuum subtraction. We also list in Table 2 the line-of-sight reddening estimates $E(B-V)$ from Schlegel, Finkbeiner & Davis (1998), as given in the NASA/IPAC Extragalactic Database, along with the corresponding typical and maximum interstellar polarizations estimated using the empirical relationships of Serkowski, Mathewson & Ford (1975). The measured polarization is significantly higher than the estimated maximum iterstellar polarization in each object.

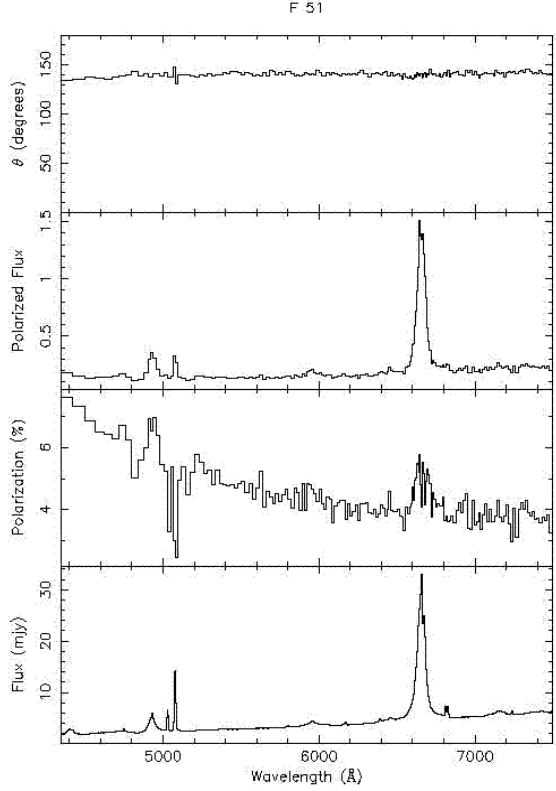

**Figure 1.** Spectropolarimetric data for Fairall 51. The panels show, from top to bottom, the position angle of polarization ($\theta$), the polarized flux density ($p \times F_\lambda$), the percentage polarization ($p$), and the total flux density ($F_\lambda$). The polarization data are binned to a constant error

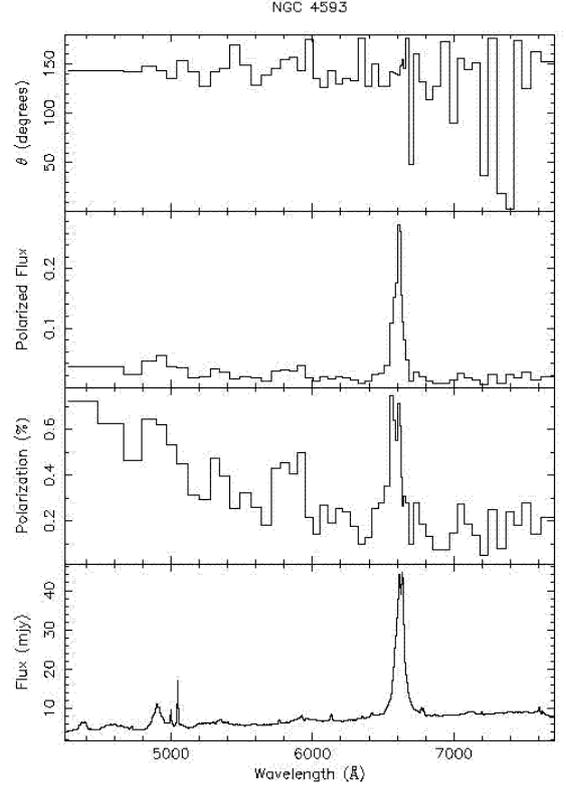

**Figure 2.** As Fig. 1 for NGC 4593. Polarization data binned to 0.1 per cent.

## 2.2 Seyfert 1 galaxies with Seyfert 2-like polarization spectra

### 2.2.1 Fairall 51

Seyfert 2-like polarization characteristics are most strikingly apparent in Fairall 51 (Fig. 1). The continuum is polarized at a level ≈4 per cent in the red (~7000Å) but $p(\lambda)$ rises steeply for wavelengths below 6000Å, to nearly 8 per cent at the extreme blue end of the spectrum. In addition, there are local increases in $p$ associated with the broad H$\alpha$ and H$\beta$ lines. In contrast, the [OIII]$\lambda\lambda$4959, 5007 lines clearly depolarize the underlying continuum and similar depolarizations in the broad H$\alpha$ line are due to [NII]$\lambda$6583 and the narrow component of H$\alpha$ itself. The spectrum is much redder in total flux than in polarized flux. In contrast to most of the intrinsically polarized Seyfert 1 galaxies in our sample, there are no significant variations in $\theta(\lambda)$ over the broad H$\alpha$ or H$\beta$ lines. To within a few degrees, the polarization PA is constant over the entire observed spectral range. The optical polarization of Fairall 51 has recently been discussed in detail by Schmid et al. (2001), whose data show very similar features. An extensive literature search yielded no published information on the radio source morphology.

### 2.2.2 NGC 4593

Compared with Fairall 51, NGC 4593 is relatively weakly polarized. Nevertheless, its polarization spectrum is clearly similar in form to that of Fairall 51 (Fig. 2). Although the continuum polarization is barely measurable in the red, it increases steeply below 6000Å, to 0.7 per cent. Again, there are strong local increases in $p$ associated with broad H$\alpha$, and less clearly, H$\beta$. A similar feature, possibly associated with the broad HeI$\lambda$5876 emission-line, is sensitive to the error binning (see S02) and may be spurious. There are no significant variations in $\theta$ over the spectrum. In polarized flux, the broad H$\alpha$ line appears slightly blue-shifted relative to the total flux line profile. The radio source is unresolved in several sets of VLA observations obtained at various frequencies

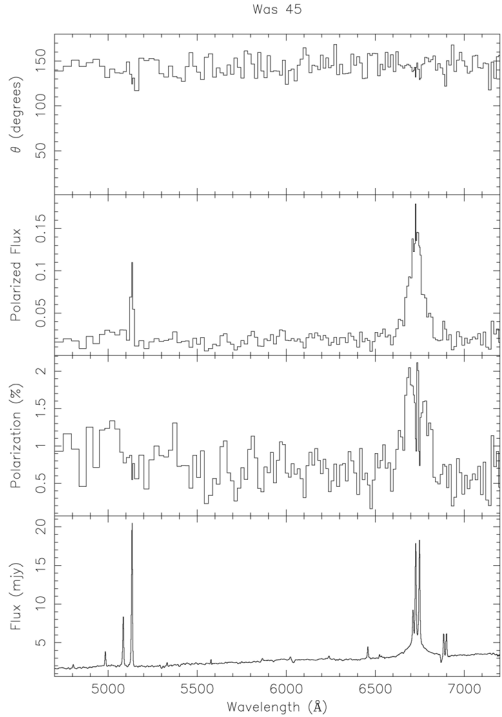
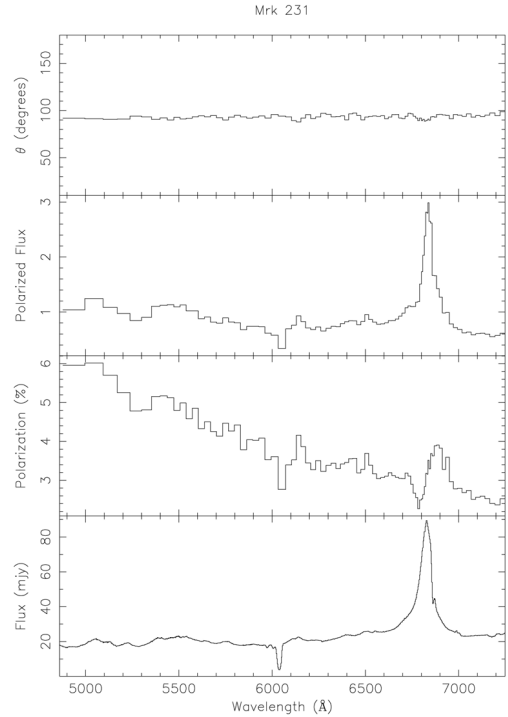

**Figure 3.** As Fig. 1 for Was 45. Polarization data binned to 0.15 per cent.

**Figure 4.** As Fig. 1 for Mrk 231. Polarization data binned to 0.2 per cent.

(Ulvestad & Wilson 1984a, 1989; Thean et al. 2000; Schmitt et al. 2001a).

### 2.2.3   Was 45 (UGC 7064)

The broad component of Hα is relatively weak in Was 45 and accordingly, this object is classified as a Seyfert 1.8 (Bothun et al. 1989). Optical spectropolarimetry has previously been obtained by Goodrich (1989a), who concluded that the observed polarization was consistent with foreground interstellar polarization. The data presented in S02 show evidence for both an increase in $p$ and a small swing in $\theta$ across the broad Hα line, indicating that the nucleus is, in fact, intrinsically polarized. However, these data have relatively low signal-to-noise and only cover the red end of the spectrum. Our new spectropolarimetry (Fig. 3) confirms the increase in $p$ over the Hα line. The adjacent continuum is polarized at a level of ≈0.6 per cent, but as in Fairall 51 and NGC 4593, there is a local increase to ≈1.5 per cent over the line itself. As a result, the broad component of Hα is much more prominent in the polarized flux spectrum than it is in total flux. There is a hint of a similar increase in $p$ across the broad Hβ line. However, unlike Fairall 51 and NGC 4593, the continuum polarization does not increase significantly to shorter wavelengths. The polarization PA is constant over the spectrum at $\theta \approx 149°$; the small rotation across Hα is not confirmed by the new data. The VLA map presented by Condon et al. (1998) shows two sources ~1 arcmin apart, which appear to be associated with Was 45 and another nearby *IRAS* source.

### 2.2.4   Mrk 231

The continuum is polarized at a level ≈2 per cent in the red (~7500Å) but $p(\lambda)$ rises to shorter wavelengths, reaching ≈6 per cent at ~5000Å and also exhibits variations across the broad Hα line (Fig. 4). Relative to the adjacent continuum, $p$ exhibits a peak in the red wing but a dip in the blue wing. The broad Hα line therefore appears redshifted in the polarized flux spectrum. The polarization position angle is relatively constant over the entire observed spectral range, with an average value $\theta \approx 95°$. VLBA observations of the nucleus (Ulvestad et al. 1999) have revealed a triple radio source ~2 pc in extent, in PA~5°. This structure is, therefore, perpendicular to the optical polarization PA. The central component of the

### 2.2.5 NGC 3227

This nearby Seyfert 1.5 is known to have both a compact (0.4 arcsec) double radio source in PA≈170°, and high-excitation line emission extending ~7 arcsec NE of the nucleus, in PA~30° (Mundell et al. 1995). It was included in the pioneering study of Schmidt & Miller (1985), whose low resolution spectropolarimetry shows $p(\lambda)$ rising from ≈1 per cent at 6000Å to ≈3 per cent at 3500Å, while the position angle is constant at $\theta$≈135°. Schmidt & Miller also reported that the Hβ line is polarized at the same level as the continuum whereas the [OIII]λλ4959, 5007 doublet has a slightly lower polarization, but the same PA as the continuum. Our new spectropolarimetry (Fig. 5) also shows an increase in the degree of polarization to the blue, from ≈0.5 per cent at 7000 Å to 1.5 per cent at 5000Å, in good agreement with Schmidt & Miller's results. However, the higher spectral resolution of our data allows better measurements of the emission line polarization, and reveals some differences. We confirm that the narrow lines are polarized at a lower level than the continuum. Sharp dips in $p(\lambda)$ are associated with the narrow component of Hα and [NII]λ6583, as well as the strong [OIII]λλ4959, 5007 lines. We also detect small changes in $\theta$, showing that these lines are polarized at a slightly different PA from that of the continuum. They are also prominent in the polarized flux spectrum. The most important new result is the clear peak in $p(\lambda)$ associated with the broad wings of Hα (Schmidt & Miller's spectropolarimetry does not extend to Hα). There is some evidence for a similar feature in Hβ, although with the lower signal-to-noise at the blue end of the spectrum it is harder to distinguish from the rising continuum polarization. Except for the small changes associated with the narrow lines and a small rotation (~10°) over the broad Hα line, the polarization PA is constant across the spectrum with a value $\theta$≈122°. It is clear from our new data that the optical polarization properties of NGC 3227 are consistent with polar scattering. In this case, there is also a well-defined radio axis, but the polarization PA is neither aligned with the radio PA, nor perpendicular to it. However, it is nearly perpendicular to the symmetry axis of the extended emission-line region. We will consider the implications of these alignments in Section 5.3.

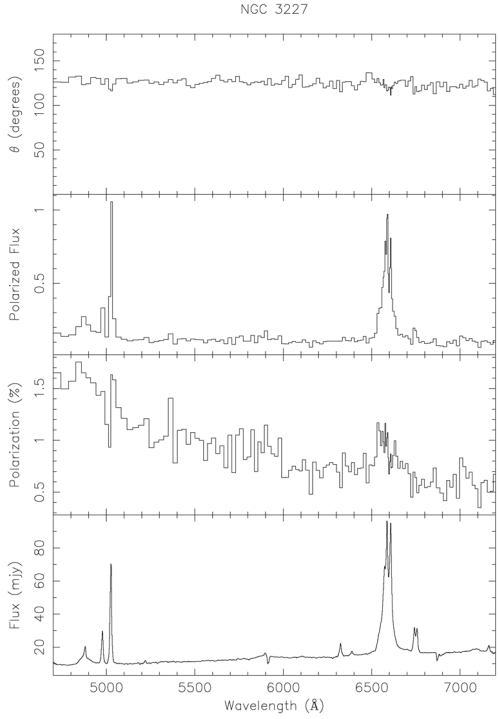

**Figure 5.** As Fig. 1 for NGC 3227. Polarization data binned to 0.08 per cent.

triple radio source also resolves into two components oriented in a roughly East-West direction. However, in view of the uncertain nature of this feature we assume the parsec-scale triple represents the main jet axis. The optical polarization of Mrk 231 has previously been discussed by GM94, whose data show similar features (however, they also note small changes in $\theta$ of 10° in the blue wing of the broad Hα line and over the He I 5876 line). These authors conclude that the observed optical polarization in Mrk 231 is due to scattering in an outflow along the radio axis. Smith et al. (1995) present a detailed study of the UV-optical polarization of Mrk 231 discussing, in particular, the wavelength dependent contribution of diluting starlight. They show that the continuum polarization continues to increase into the near UV, reaching a maximum between 2755 and 2797Å, where $p\sim18$ per cent. At shorter wavelengths, the polarization declines dramatically and Smith et al. propose that this is due to dilution of the polarized flux by the continuum produced by a population of hot OB stars. The drop in $p$ is coupled with a 40° PA rotation, the origin of which, however, is unclear.

*2.2.6    ESO 323–G077*

Spectropolarimetry of this bright Seyfert 1 has recently been obtained with the VLT by Schmid, Appenzeller & Burch (2003). The observed continuum polarization increases from $p\approx2$ per cent at 8300 Å, to $p\approx8$ per cent at 3600 Å. There are also local increases in $p$ associated with various broad emission lines including H$\alpha$, H$\beta$ and H$\gamma$, He I$\lambda$5876 and the Fe II features centred at 5200 and 4750Å. The narrow lines are unpolarized and sharp dips in polarization caused by dilution by the strong [O III]$\lambda\lambda$4959, 5007 lines are clearly visible. The polarized flux spectrum is bluer than the total flux spectrum. The polarization PA is independent of wavelength, except for an apparent small (<10°) swing associated with the broad H$\alpha$ line, which, though smaller in amplitude, has the same general form as the PA swings often seen in equatorially-scattered Seyfert 1 galaxies. The radio source has been studied by Nagar et al. (1999) whose maps are, however, too noisy to determine a reliable PA. Narrow band imaging by Mulchaey, Wilson & Tsvetanov (1996) reveals [OIII] emission extending South of the nucleus over several kiloparsecs, presumably tracing an ionisation cone. As noted by Schmid et al., this is nearly perpendicular to the average polarization PA ($\theta$=84±2°).

*2.2.7    Mrk 704*

Spectropolarimetric observations of Mrk 704 obtained by GM94 and M96 show features very similar to those seen in Fairall 51. The continuum polarization rises to the blue, increasing from ~2 percent at 7500Å to ~6 per cent at 3500Å. There are local peaks in $p$ associated with the cores of the broad Balmer emission lines but in addition, the degree of polarization dips slightly below that of the continuum in the broad-line wings (see fig. 5 and table 2 in GM94). This behaviour is, in fact, consistent with polar scattering of line emission from a rotating disc (the line profile in polarized flux is narrower than that in total flux profile, leading to greater dilution of the percentage polarization in the line wings) but the exact opposite of that expected for the equatorial scattering geometry (S02). The polarization PA is roughly constant across the spectrum, at $\theta\approx70°$. There is a small rotation of $\theta$ across H$\alpha$, but it is not clear if this is real or due to an instrumental effect. Thean et al. (2000) classify the radio source as unresolved, although their map, obtained at 8.4 GHz with the VLA in A-configuration, appears to show two weak components in PA~135°.

*2.2.8    Mrk 376*

Mrk 376 was also observed by GM94 and is generally quite similar to Mrk 704. The continuum polarization increases from ~3 per cent at 7000Å to ~5 per cent at 3500Å. There are indications of local increases in $p$ associated with the red wings of the broad H$\alpha$ and H$\beta$ lines, suggesting that the lines are red-shifted in polarized light, but these features are less prominent than in Mrk 704 and most other sources. The polarization PA is constant over the spectrum, except for an apparent small rotation ~10° in the blue wing of H$\alpha$. Papadopoulus et al. (1995) report that the radio source is unresolved in an image obtained at 6 cm with the VLA in B-configuration.

*2.2.9    Mrk 766*

Mrk 766 was observed by Goodrich (1989b) as part of a study of the optical polarization of narrow-line Seyfert 1 (NLS1) galaxies. In this object, the continuum polarization again increases strongly to the blue, from ~1.5 per cent at 7000 Å to ~4.5 per cent at 4000 Å. Local increases in $p$ associated with the broad H$\alpha$ and H$\beta$ lines are present, as are sharp dips corresponding to the [O III]$\lambda\lambda$4959, 5007 lines. The polarization PA is reported to be independent of wavelength, at $\theta$=90.0±0.3°. Radio observations have been obtained by several different groups (Ulvestad & Wilson 1984b; Ulvestad, Antonucci & Goodrich 1995; Kukula et al. 1995; Nagar et al. 1999; Thean et al. 2001), all of whom consider the radio source to be slightly resolved, with estimates for the PA ranging from 12 to 34°. We follow Nagar et al. in adopting PA=27° as representative of the source axis on the smallest scales. Mulchaey et al. (1996) report [OIII] emission slightly extended in the NW–SE axis but it is not clear if this traces an ionization cone.

*2.2.10    IRAS 15091–2107*

Another of the NLS1s observed by Goodrich (1989b), the polarization properties of this object are very similar to those of Mrk 766. The continuum polarization increases from ~3 per cent at 7000 Å to ~6.5 per cent at 4500 Å. The degree of polarization increases over the broad H$\alpha$ and H$\beta$ lines, although these features are less prominent than in Mrk 766. The [O III]$\lambda\lambda$4959, 5007 lines are again associated with local minima in the continuum polarization. The polarization PA is constant ($\theta$=61.7±0.2°). Condon et al. (1998) and Ulvestad et al. (1995) consider the radio source to be unresolved in their VLA maps but Thean et al. (2000) describe it as slightly resolved in PA ~2°, based on an 8.4 GHz map obtained with the VLA in A-configuration.

### 2.2.11 Mrk 1239

The polarization spectrum of Mrk 1239, also part of Goodrich's (1989b) study, is somewhat more complex. The continuum polarization shows the familiar wavelength dependence with $p(\lambda)$ rising from ≈2 per cent 7000 Å to ≈4.5 per cent at 4500 Å, while $\theta(\lambda)$ is approximately constant at ≈130° over this range. There is an increase in polarization over the broad H$\alpha$ and H$\beta$ lines, but in both cases, this is confined to the red wing (there is a hint of a slight decrease in the blue wings). In polarized flux, therefore, the broad Balmer lines exhibit extended red wings, which are not visible in total flux. A small change in $\theta$ (~10°) appears to be associated with the polarization peak in the red wing. The [OIII]$\lambda\lambda$4959, 5007 lines are also polarized, but at a lower level and a different position angle. In general, therefore, the polarization properties of Mrk 1239 are similar to those of Mrk 231. Although Mrk 1239 has a relatively strong radio source, it is unresolved in published VLA maps (Ulvestad et al. 1995; Nagar et al. 1999; Thean et al. 2000). Mulchaey et al. (1996) find [OIII] emission slightly extended in the NE–SW direction.

### 2.2.12 Mrk 1218 (NGC 2622)

Originally classified as a Seyfert 1.8 (Osterbrock & Dahari 1983), the broad Balmer lines in Mrk 1218 underwent a large increase in flux between 1981 and 1986, sufficient to change its spectrum to that of a Seyfert 1. Spectropolarimetric observations were obtained by Goodrich (1989a) in 1986 and 1987 during this state. In these data, $p(\lambda)$ increases from ≈1.5 per cent at 7000 Å to ≈4.5 per cent at 4000 Å. Local peaks in $p$ are associated with the broad H$\alpha$ and H$\beta$ lines, but appear to be slightly blueshifted relative to the total flux line profiles (in contrast to the redshifts seen in Mrk 231 and Mrk 1239). The polarization PA is approximately constant across the spectrum at $\theta$~65°, except in the red wing of H$\alpha$ where it rotates by about 10°. There does not appear to be a matching rotation in H$\beta$. Nagar & Wilson (1999), citing VLA observations by Ulvestad (1986), note that the radio source is slightly extended in PA 155°. However, we have been unable to find any reference to the quoted PA in the primary source.

### 2.3 General comments

Of the 20 intrinsically polarized Seyfert 1 galaxies in our original sample, only three (Fairall 51, NGC 4593 and Was 45) exhibit spectropolarimetric properties consistent with polar scattering. However, through a literature search and new observations, we have now identified 12 such objects. These all show very similar polarization characteristics. The degree of polarization increases strongly to the blue and has local maxima associated with the broad Balmer lines. The polarization PA is generally constant over the spectrum. All these features are characteristic of PBL Seyfert 2 galaxies, in which the polarization is generally believed to be due to far-field scattering of light from the obscured AGN by dust or electrons located along the polar axis of the torus. Two objects, Mrk 231 and Mrk 1239, show more complex polarization variations over the broad H$\alpha$ profile, with a minimum in $p$ in the blue wing and a peak in the red wing. This behaviour, however, can easily be understood if the scattering medium is participating in a radial outflow, away from the nucleus. In this case, the scattered line profile will be redshifted relative to that in direct light and differential dilution produces the observed trough-peak structure. Some objects show changes in $\theta$ across the broad-lines, but these are small compared to the large blue-red swings seen in the 'equatorially-scattered' Seyfert 1 galaxies. It is possible that these small PA swings are due to a minor contribution to the polarized flux from a different scattering route, perhaps even the equatorial route (Section 4).

The logic of the Seyfert unification scheme demands that this polar scattering route should be present in all Seyfert nuclei. Fairall 51 and similar objects therefore appear to represent a group of type 1 nuclei in which, as in type 2's, polar scattering (rather than equatorial scattering) dominates the observed polarization. This being so, the polarization PA in these Seyfert 1 galaxies should be perpendicular to both the projected radio source axis and also that of any emission line 'ionization cone' due to the escape of ionizing radiation along the poles of the torus. This, as already noted, is almost invariably the case in Seyfert 2 galaxies. Unfortunately, this prediction is not easy to verify for most of our objects. Although most have radio observations, the radio source is either unresolved or marginally resolved in many cases, making it difficult to determine the radio axis PA unambiguously. In only two cases, Mrk 231 and NGC 3227, is there a reasonably well-defined jet axis. In the former, the polarization PA is roughly perpendicular to the jet axis. In the latter, it is offset by 42° from the radio axis, but nearly perpendicular to the major axis of the ionization cone. Similarly, the polarization PA is perpendicular to the ionisation cone in ESO 323−G077. There are two more objects, Mrk 766 and *IRAS* 15091−2107, which are reported to have slightly extended radio sources and in these cases the polarization PA is within about 30° of the perpendicular to the radio PA. In all cases in which the PA of either the radio source or ionization cone can be determined, the polarization PA is approximately perpendicular to one or the other (i.e. has an offset ≥60°). Therefore, the orientation of the electric vector (**E**) of the scattered light relative to the projection on the sky of the torus axis (insofar as we can reliably determine it) is consistent with polar scattering. We will consider the relative orientations of the polarization PA and the radio axis further in Section 5.3.

## 3   THE POLARIZATION SPECTRUM OF SEYFERT 2 NUCLEI

As the polarization spectra of the polar-scattered Seyfert 1 galaxies bear a strong resemblance to those of PBL Seyfert 2 galaxies, it is worth briefly outlining the mechanisms that are thought to operate in the latter to produce their characteristic polarization signature.

As already noted, the presence of polarized broad-lines in Seyfert 2 galaxies is due to scattering of light escaping from the AGN hidden within the circum-nuclear torus. In the simplest picture, light from the BLR and central continuum source emerges in bi-polar 'radiation cones' which illuminate the narrow-line region (NLR) outside the torus (the counter-cone may be obscured by dust in the disc of the host galaxy or in the torus). Free electrons and/or dust associated with the NLR scatter some of the light into our line-of-sight, and hence cause it to become polarized. Since light polarized by scattering has an **E** vector perpendicular to the scattering plane (the plane containing the incident and scattered ray), the scattered light has a polarization PA perpendicular to the photons' projected direction of flight prior to scattering. Given that the principal axis of the active nucleus is defined by the rotation axis of the accretion disc and that the axes of the radio source and the torus are co-aligned with this axis, the scattered light (averaged over the scattering cone) will be polarized perpendicular to the projection of the radio source axis on the sky. This basic picture explains why some Seyfert 2 galaxies exhibit type 1 spectra in polarized light and also accounts for the generally observed perpendicular orientation of the polarization **E** vector relative to the radio source. However, some refinement is necessary in order to explain the observed polarization spectrum in detail.

We would expect light from a pure polar-scattered source to be uniformly polarized at a level of a few tens of per cent (given the large scattering angles that are implied for Seyfert 2 galaxies). In fact, optical polarizations of only a few per cent are typically observed (NGC 1068 itself is an exception). Similarly, since the broad-line and continuum photons follow the same scattering path, light from these two sources should be identically polarized. In fact, local peaks in *p* are often observed at the wavelengths of the broad-lines. None the less, there is usually little change in $\theta$, which suggests that the line and continuum photons do indeed follow similar scattering paths. These effects can be explained if the polar-scattered flux is diluted by unpolarized continuum emission. Whilst this will decrease *p* for both the scattered line and continuum emission, the larger scattered flux in the broad-lines will suffer relatively less dilution, resulting in local increases in *p* over the lines (e.g. see the discussions in Antonucci 1993 and Tran 1995). Although the exact source of this diluting unpolarized continuum has been widely debated (e.g. Tran 1995 and references therein), it now seems clear that it is stellar in origin (Antonucci 1993, Heckman et al. 1995; Gonzalez-Delgado et al. 1998; Storchi-Bergmann, Cid-Fernandes & Schmitt 1998).

The increase in $p(\lambda)$ towards shorter wavelengths observed in many Seyfert 2 nuclei can also be explained by the presence of a diluting continuum component. Raleigh scattering by small dust grains increases strongly to shorter wavelengths but although the scattered spectrum is significantly bluer than the incident spectrum, the *percentage* polarization is wavelength independent unless the scattered flux is diluted by a source of unpolarized light. On the other hand, electron scattering is independent of wavelength and in this case, the diluting component must be red (e.g. an old stellar population) in order to produce an increase in $p(\lambda)$ to the blue. Both electron and dust scattering are likely to contribute to the polarization spectrum of Seyfert nuclei at some level. This is the case in NGC 1068 (e.g. Miller, Goodrich & Matthews 1991; Inglis et al. 1995), where electron scattering dominates close to the nucleus whilst dust scattering occurs in more distant clouds. In the electron-scattered light, the polarized broad-lines are broader than they are in the dust-scattered light, implying a thermal temperature for the scattering electrons of ~300,000 K. In most other, more distant, Seyfert 2 galaxies, the electron- and dust-scattering regions cannot be resolved, and therefore the observed polarization spectrum is almost certainly the result of scattering by a mixture of electrons and dust.

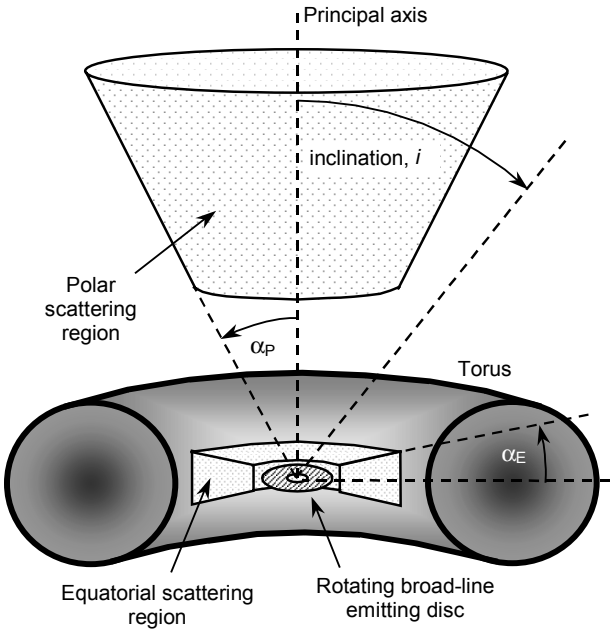

**Figure 6.** Two-component scattering geometry proposed to explain the optical polarization characteristics of Seyfert nuclei. A rotating disc emits the broad Balmer lines. The equatorial scattering region is modelled as a flared disc of half-opening angle, $\alpha_E$, and closely surrounds the emission disc. The polar scattering region is modelled as a truncated cone of half-opening angle $\alpha_P$, aligned with the axis of the circum-nuclear torus. We assume that the symmetry axes of the emission disc, both scattering regions and torus are co-aligned and define the principal axis of the system. Inclination, *i*, is measured from the principal axis to the line-of-sight.

In general, therefore, the optical polarization spectrum of a type 2 Seyfert nucleus is the result of a mixture of several components.

i. Direct light from the narrow-line region, including both line emission and the nebular continuum (mainly recombination and thermal Bremsstrahlung emission);
ii. Light from the broad-line region and central continuum source of the obscured AGN, scattered and hence polarized by electrons and/or dust distributed within the torus illumination cone, which is aligned with the radio axis;
iii. Starlight from the host galaxy, possibly including both young and old populations.

The contribution of each component is wavelength-dependent and likely to vary from object to object. In addition, the intrinsic spectrum is likely to be modified by interstellar dust along the line-of-sight in the both the host galaxy and our own. This will result in both wavelength-dependent extinction and, if the dust grains are aligned to a significant degree, polarization due to dichroic absorption.

The unification scheme leads us to expect that all the above components should be present even when the inclination is such that we see directly into the torus – that is, when the source appears as a Seyfert 1 nucleus. In this case, the direct light spectrum will be dominated by BLR and continuum emission from the active nucleus. The spectrum in polarized light will presumably have the additional contribution of scattered AGN light from within the torus. As we have already discussed, the observational evidence indicates that whilst equatorial scattering within the torus dominates in many Seyfert 1 galaxies, the 'Seyfert 2' polar scattering route dominates in others. The question we must now consider is, what is it that determines which of the two sources of scattered light dominates the observed polarization of Seyfert 1 nuclei?

## 4 THE SCATTERING GEOMETRY OF TYPE 1 SEYFERT NUCLEI

We argued in S02 that much of the observed diversity in the optical polarization properties of Seyfert nuclei is well-explained by a model in which the broad-line emission originates from a rotating disc and is scattered by two distinct scattering regions that produce orthogonally polarized light. These are:

i. a scattering region co-planar with the line-emitting disc and situated within the torus, in its equatorial plane – the 'equatorial' scattering region;

ii.   a scattering cone situated outside the torus but aligned with the torus/emission disc axis − the well established 'polar' scattering region.

The proposed scattering geometry is shown in Fig. 6. Detailed modelling of the wavelength dependence of polarization ($p(\lambda)$ and $\theta(\lambda)$) over the broad H$\alpha$ line resulting from the equatorial scattering region is presented elsewhere (Smith 2002; S04).

Consideration of the orientation of the polarization PA relative to the radio axis suggests that equatorial scattering tends to dominate the observed polarization in Seyfert 1 galaxies, whereas polar scattering dominates in Seyfert 2 galaxies. However, as we have seen, this apparent separation of dominant scattering route by Seyfert type is not clear-cut; a significant number of Seyfert 1 galaxies exhibit polarization properties most readily explained by polar, not equatorial, scattering.

For polar scattering to dominate, we simply require that the observer receive more polarized flux from the polar than from the equatorial scattering region. The polar scattering region is visible in both Seyfert 1 and 2 galaxies since it is located outside the circum-nuclear torus. Polarized flux from this component will therefore contribute to the observed net polarization in *all* Seyfert nuclei. This component must dominate in Seyfert 2 galaxies, because the torus blocks the direct view of the continuum source, the BLR *and* the equatorial scattering region. Conversely, the observational evidence suggests that when the direct view of the nuclear regions is not blocked by the torus, equatorial scattering tends to dominate the observed polarization. Why is this not the case in the polar-scattered Seyfert 1 galaxies, in which we evidently have a direct view of the BLR? One possibility is that the equatorial component is unusually weak in these objects, allowing the polar component to dominate. This could occur if, for example, the column density of scatterers is relatively small in the equatorial region, compared to the polar region. However, a more interesting possibility, in the context of the unification scheme, is that geometry plays a role in suppressing the equatorial component relative to the polar component. There are two geometrical mechanisms that can potentially lead to a greater amount of polarized flux being received from the polar region than from the equatorial region:

i.   as the degree of polarization is dependent on scattering angle, polar scattering may be favoured over equatorial scattering for certain orientations of the principal axis of the AGN;
ii.   the system axis may be oriented such that the line-of-sight to the nucleus passes through the 'top edge' of the torus, resulting in partial extinction of scattered flux from the equatorial region.

In sections 4.1 and 4.2 we discuss these two mechanisms in turn, with the aid of the Generic Scattering Model (GSM) for AGN described by Young (2000).

### 4.1   The inclination dependence of polarization due to polar and equatorial scattering

We assume that the rotation axis of the line emitting disc and the polar axis of the torus are co-aligned and define the principal symmetry axis of the AGN. The average scattering angle (that between the incident and scattered rays) for the polar and equatorial scattering regions will in general be different at any given inclination of this axis to the observer's line-of-sight. As the degree of polarization of scattered light is a strong function of the scattering angle (e.g. van der Hulst 1957) it seems reasonable to expect that the contributions of the two scattering regions to the net polarization might vary in such a way that they dominate over different ranges in inclination.

We have used the GSM to investigate this possibility. The polar scattering region is represented as a cone whilst the equatorial scattering region (referred to as the compact scattering region in Young (2000)) is modelled as a flared disc (Fig. 6). Both the cone and the disc are separately specified by half-opening angles and inner and outer radii. In the models considered here, both structures have their symmetry axes aligned with the principal axis. The scattering particles in both regions are assumed to be free electrons whose number density has a radial dependence given by a power law. The GSM implicitly assumes that the scattering regions are optically thin, so that only single scattering need be considered. This is probably a good approximation for the polar scattering region but is less secure for the equatorial scattering region, whose properties (apart from its basic geometry) are unknown. Nevertheless, simulations by Henney & Axon (1995) suggest that even for optically thick scattering media, the polarization is dominated by first order scattering and therefore, the resultant polarization signature approximates to that for single scattering.

**Table 3.** Model parameters for the polar and equatorial scattering regions.

| | |
|---|---|
| Polar scattering region | |
| inner radius | $3.0 \times 10^{16}$ m |
| outer radius | $1.0 \times 10^{19}$ m |
| number density at inner radius | $1.0 \times 10^{11}$ m$^{-3}$ |
| radial dependence of number density | $r^{-2}$ |
| Equatorial scattering region | |
| inner radius | $1.0 \times 10^{15}$ m |
| outer radius | $1.0 \times 10^{16}$ m |
| number density at inner radius | $1.5 \times 10^{12}$ m$^{-3}$ |
| radial dependence of number density | $r^{-1}$ |

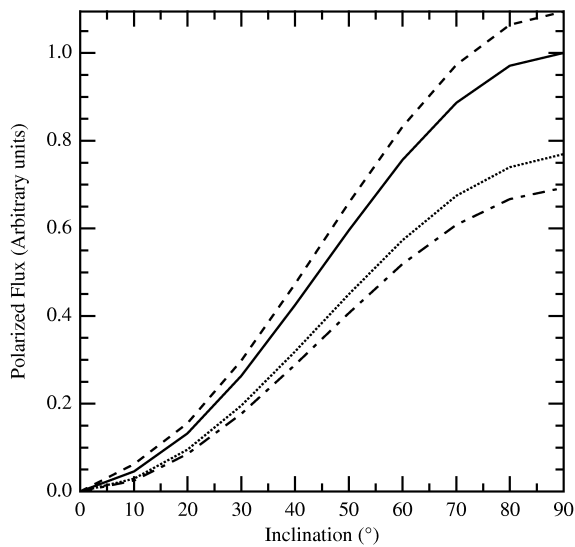

**Figure 7.** Variation of polarized flux with inclination for the polar scattering region only. Fluxes for four values of the half-opening angle of the polar scattering cone ($\alpha_P$; c.f. Fig. 6) are shown: 35° (dotted line), 45° (full), 55° (dashed line), 75° (dot-dashed). The fluxes are normalized to the value at $\alpha_P$=45°, $i$=90°.

The GSM was used to calculate the polarized flux produced by both equatorial and polar scattering regions as a function of the inclination ($i$) of the principal axis to the observer's line-of-sight. We consider both regions to be spatially unresolved and therefore the polarized flux is that obtained by integrating over all spatial elements. With the exception of the half-opening angles for the polar scattering cone and the equatorial scattering disc, parameter values were taken from our detailed model of Mrk 509 (Young 2000), which accurately reproduces the observed $p(\lambda)$ over the broad H$\alpha$ emission-line. We are here only interested in the total polarized flux, and not in its wavelength dependence. The relevant parameters are those listed in Table 3.

### 4.1.1 Individual components

We will first consider the polar and equatorial scattering regions separately. The variation of polarized flux with inclination is shown in Fig. 7 for the polar scattering cone and in Fig. 8 for the equatorial scattering disc. In each case, results are shown for 4 values of the half-opening angle: 35, 45, 55 and 75° for the polar and 10, 20, 30 and 50° for the equatorial region, respectively. The inclination was varied from 0° (principal axis aligned with line-of-sight) to 90° (principal axis orthogonal to line-of-sight) in 10° steps. The plotted curves show similar behaviour for both the polar and equatorial scattering regions. The polarized flux increases monotonically from zero at $i$=0 to a maximum at $i$=90°, as might be expected for Rayleigh scattering.

However, although the curves for the two regions are rather similar, their behaviour is, in fact, the result of different effects. When the principal axis is aligned with the line-of-sight ($i$=0°) both scattering regions exhibit circular symmetry on the sky plane. There is complete cancellation of the polarization vectors and thus null polarization. Once the system axis is tilted with respect to the line-of-sight, the circular symmetry is broken, resulting in a non-zero net polarization. In the case of the polar scattering cone, the increase in the degree of polarization of the scattered light is due simply to the fact that the average scattering angle increases with $i$. The behaviour of the equatorial disc is rather more complex. In this case, when viewed face-on ($i$=0), the scattering angle for all points in the disc is (depending on the degree of flare) close to 90°. Therefore, the polarization for each individual ray approaches 100 per cent, but as already noted, cancellation ensures null polarization for the disc as a whole. As $i$ is increased, the scattering angle remains at ~90° for scatterers located along the major axis of the inclined disc, whereas it decreases for those located along the orthogonal (minor) axis (see discussions in Smith 2002; S04). As a result, cancellation around the disc becomes less complete as $i$ increases, resulting in an increase in the net polarization, and hence in the polarized flux.

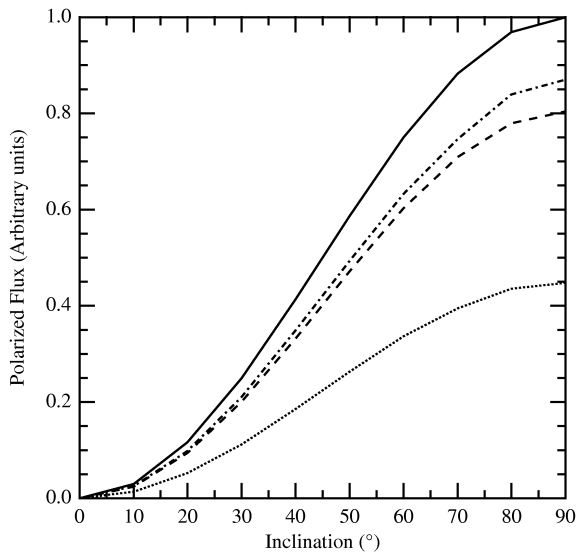

**Figure 8.** The variation of polarized flux with inclination for the equatorial scattering region only. Fluxes for four values of the half-opening angle of the flared disc ($\alpha_E$; c.f. Fig. 6) are shown: 10° (dotted line), 20° (dashed), 30° (full), 50° (dot-dashed). The fluxes are normalized to the value at $\alpha_E=30°$, $i=90°$.

For a given (non-zero) inclination, the polarized flux from both polar and equatorial scattering regions increases with the half-opening angle of the scattering region up to a certain critical angle, beyond which the polarized flux decreases. The initial increase is simply a result of the increased coverage of the sky by the scattering region, as seen by the emitting source. However, when the half-opening angle is 45°, cancellation between orthogonally polarized rays starts to occur and at some point beyond 45°, this effect becomes strong enough to cause a reduction in the net polarization despite the increased sky coverage. In the limiting case when the half-opening angle is 90° (when the polar and equatorial scattering regions are hemispherical and complete spherical shells, respectively) each ray has an orthogonally polarized counterpart, resulting in zero net polarization.

### 4.1.2 Combined polar and equatorial scattering

In our two-component scattering model, the net polarization of the light emitted by the Seyfert nucleus is produced by a combination of polar and equatorial scattering. In this section, we discuss the inclination dependence of the total polarized flux resulting from this scattering geometry, ignoring, for the moment, obscuration by the torus.

Fig. 9 shows the variation of polarized flux with inclination for both a polar scattering cone and an equatorial scattering disc. The half-opening angles of the cone and disc scattering regions are 45 and 30°, respectively, the other parameters having the values listed in Table 3. With this fairly arbitrary choice of parameters, the equatorial scattering region produces the largest polarized flux. This situation could easily be reversed with different choices of half-opening angles or number densities for either component, but the observational evidence described in S02 suggests that equatorial scattering does indeed dominate in most Seyfert 1 galaxies. The total polarized flux resulting from the combination of the fluxes from the two scattering regions is also shown in Fig. 9. When the system is viewed pole-on, both scattering regions are circularly symmetric in projection and cancellation ensures that the polarization of the scattered light is zero. As $i$ is increased, the total flux, not surprisingly, shows behaviour similar to that of the individual scattering regions, also reaching a maximum at $i=90°$. However, the total polarized flux is always less than that produced by the dominant scattering region, since equatorial and polar scattering produce orthogonal polarizations, resulting in partial cancellation when the fluxes are combined. The net polarization position angle will be that of the dominant scattering route – in this case, since equatorial scattering dominates, parallel to the principal axis.

The monotonic nature of the polarized flux–inclination curves for the equatorial and polar scattering regions, together with the fact that they are both anchored at the origin (since the polarized flux must vanish at $i=0$), implies that they cannot intersect for any value of $i$ in the range 0–90°. This is the case regardless of how they are normalised relative to each other. Thus, if the parameters governing covering factor and column density are chosen so that the polarized flux from the equatorial scattering region exceeds that from the polar scattering region at $i=90°$, this will also be true for all inclinations down to $i=0°$. It follows that we cannot explain the polar-scattered Seyfert 1 galaxies (Section 2) simply in terms of a variation in the average scattering angle with the orientation of the system axis. For example, if equatorial scattering dominates in the more face-on systems ($i \rightarrow 0°$), then other things being equal, it should also dominate in the more inclined sources.

In reality, the covering factors and scattering column densities of the polar and equatorial scattering regions are likely to differ significantly from object to object. The dominance of the polar scattering route in a minority of some Seyfert 1 galaxies may, therefore, simply be a function of the circum-nuclear environment. Nevertheless, the very close similarity between the polarization properties of the polar-scattered Seyfert 1 galaxies and PBL Seyfert 2 galaxies suggests that they result from similar effects. Therefore, in the context of the unification scheme, a more attractive possibility is that the circum-nuclear torus plays a role in determining which scattering route dominates the observed polarization.

## 4.2 Extinction through the torus atmosphere

According to the Unification Scheme, when the AGN is viewed at a large inclination to its principal axis the circum-nuclear torus intercepts the direct line-of-sight to the nucleus – that is, the central continuum source, BLR *and* the equatorial scattering region are obscured. Since only the NLR and polar scattering region remain visible we would observe a PBL Seyfert 2. An obvious possibility is that polar-scattered Seyfert 1 galaxies are objects viewed at intermediate inclinations, in which the direct line-of-sight to the AGN passes through a relatively small optical depth in the upper layers of the dusty torus. The idea that some broad-line AGN are viewed through a dusty screen is not new. Wills et al. (1992) show that the polarization and spectral properties of the infrared selected Quasi-stellar object (QSO) *IRAS 13349+2438* can be explained as a combination of a reddened spectrum transmitted through a dusty thick disc (conceptually equivalent to the torus) and a polarized component due to scattering of light escaping in the polar directions. Similarly, Schmid et al. (2001) have recently modelled the spectrum of Fairall 51 as a combination of reddened direct light and dust-scattered polarized light. They propose that the polarized light comes from a dusty polar scattering region while the direct line-of-sight to the AGN passes through part of the circum-nuclear torus and is subject to an optical depth $\tau \sim 1$.

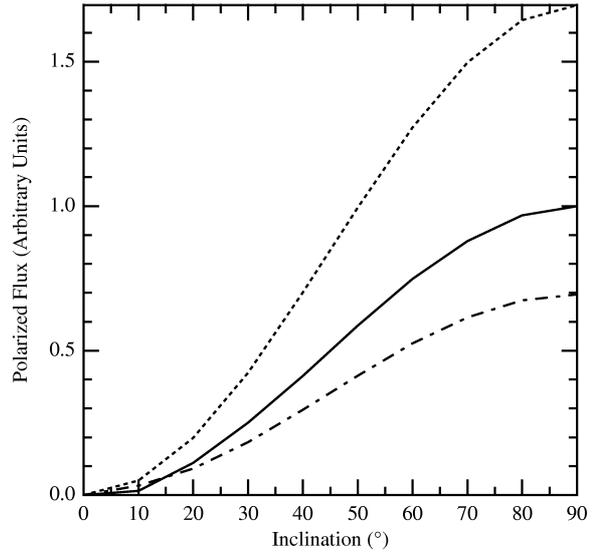

**Figure 9.** The variation of polarized flux with inclination for the two-component (polar and equatorial) scattering geometry. The polar component (dot-dashed line) has a half opening angle of 45° whilst the equatorial component (dashed line) has a half opening angle of 30°. The full line shows the net polarized flux after combining the fluxes of both individual components. The fluxes are normalized to the net flux at $i=90°$.

In our two component scattering model, a shallow line-of-sight passing through the upper layers of the torus becomes a defining characteristic of polar-scattered Seyfert 1 galaxies. Since Seyfert 1 galaxies that exhibit equatorial scattering characteristics have low polarizations (typically ~ 1 per cent), relatively little extinction along the direct line-of-sight to the nucleus would be needed to allow polar scattering to dominate the observed polarization. In order for this to occur, we require that, after suffering extinction in passing through the upper layers of the torus, the polarized flux from the equatorial scattering region is less than the unattenuated polarized flux from the polar scattering cone. The ratio of the unextincted polarized flux produced by equatorial scattering, $F_{eq}(i)$, to that produced by polar scattering, $F_{pol}(i)$, sets a lower limit to the required optical depth through the torus atmosphere: $\tau \geq \ln[F_{eq}(i)/F_{pol}(i)]$. We have used the GSM configured as in Fig. 6 to obtain estimates for $F_{eq}(i)$ and $F_{pol}(i)$. The half-opening angles of the scattering cone and equatorial scattering region were 45 and 20° respectively. Other model parameters are as given in Table 3. Statistical arguments based on the relative numbers of type 1 and type 2 Seyferts (Schmitt et al. 2001b) suggest that the torus half-opening angle is ~45°. Therefore, extinction through the upper layers of the torus is likely to become important for inclinations around 45°. For this value of $i$, we find $\tau \approx 1$, corresponding to an optical extinction $A_V \sim 1$ magnitude.

The BLR itself will be subject to the same extinction as the equatorial scattering region. However, a much larger extinction, $A_V \geq 4$, would typically be required to render the direct broad-line flux undetectable against the underlying stellar continuum. For example, assuming that the peak flux density in the broad-line is 3× that of the total continuum and that starlight contributes ~50 per cent of the latter, the line would not be detected at a signal-to-noise ratio ~10 if the extinction along the direct line-of-sight to the AGN is $A_V = 4$.

While the precise values of $A_V$ are model–dependent, the above arguments show that it is easily possible for light from the equatorial scattering region to suffer enough attenuation to allow polar scattering to dominate the observed polarization spectrum, without causing significant obscuration of the BLR and AGN continuum source.

In general, we might expect that the optical depth along the line-of-sight ($\tau_V(i)$) will increase with the inclination of the torus axis for $i \geq 45°$, as a result both of the increased path length through the torus and, presumably, an increasing dust density towards the equatorial plane. It is therefore reasonable to suppose that Seyfert 1 galaxies with polar scattering characteristics will be observed over some range in inclination governed by $\tau_V(i)$. The lower bound to this range will be the inclination at which $\tau_V(i)$ becomes large enough to attenuate the polarized flux from the equatorial scattering region to a level below that of the polar scattering region. The upper bound will be the inclination at which $\tau_V(i)$ is large enough to extinguish the direct broad-line emission relative to the stellar continuum, leaving a PBL Seyfert 2 spectrum.

The scenario outlined above predicts that polar-scattered objects should, on average, be more reddened than the general Seyfert 1 population. The extinction $A_V \approx 1$ needed to suppress equatorial scattering in the polar-scattered Seyfert 1 galaxies would produce a reddening $E(B–V) \approx 0.36$ (assuming $R_V = 3.1$, Savage & Mathis 1979). This seems to be consistent with the available observations. Winkler (1997) has determined the nuclear extinction for a large number of Seyfert 1 galaxies, including 3 objects that we have identified as polar-scattered objects. Fairall 51, ESO 323−G77 and Mrk 704 have values $E(B–V)$=0.58±0.05, 0.60±0.07 and 0.23±0.08, respectively, the first two being amongst the most reddened objects in Winkler's sample. In contrast, there is no significant extinction in Akn 120, NGC 3783 and Mrk 509, which, in our model are dominated by equatorial scattering (S02).

If the direct ray from the BLR and equatorial scattering region passes through the dusty atmosphere of the torus, we might expect dichroic absorption to cause a certain amount of polarization. For example, the line-of-sight polarization induced by the interstellar medium in our Galaxy is typically ~3×$E(B–V)$ (Serkowski et al. 1975). This would imply a dichroic polarization of ~1 per cent for the amount of reddening, $E(B–V)$=0.36, that we estimated above. This is relatively low compared to the observed polarizations of most polar-scattered Seyfert 1 galaxies, but it is comparable to those of NGC 4593, NGC 3227 and Was 45. However, the strength of any dichroically polarized flux component will depend on the magnetic field structure in the upper atmosphere of the torus. The strong polarization wavelength dependence ($p(\lambda)$ increasing to the blue) characteristic of the polar-scattered objects argues against significant dichroic polarization. It may be that the upper atmosphere of the torus is turbulent, precluding any large-scale organization of the magnetic field and hence any significant dust grain alignment.

## 5  AN ORIENTATION SEQUENCE FOR SEYFERT NUCLEI

In the context of the two-component scattering geometry, the polar-scattered Seyfert 1 galaxies are objects in which polarized flux from the polar scattering region significantly exceeds that from the equatorial scattering region. As discussed in Section 4, this situation could arise for either of two reasons:

i.  the density and/or geometry of the respective scattering regions vary from object to object in such a way that polar scattering sometimes dominates the net polarization;
ii.  polarized flux from the equatorial scattering region is attenuated by dust in the upper layers of the circum-nuclear torus.

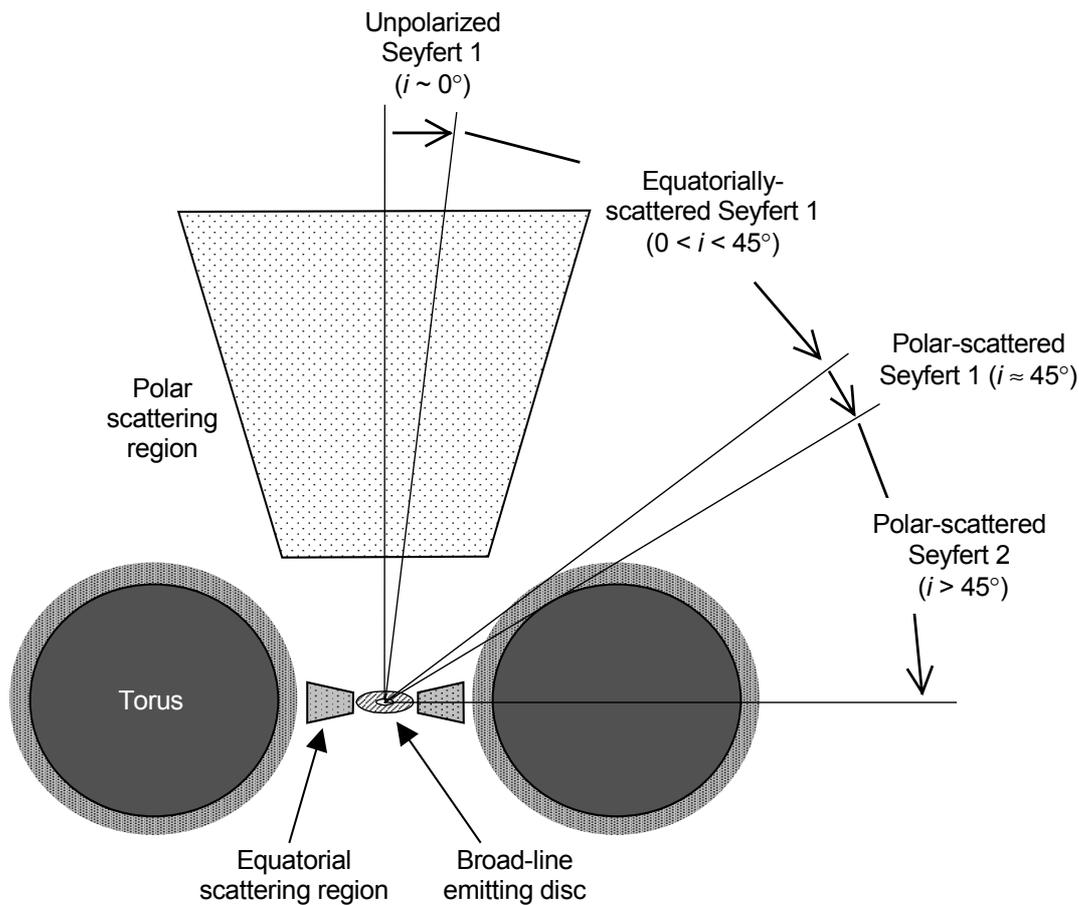

**Figure 10.** Diagram illustrating how the broad Hα polarization characteristics of Seyfert galaxies can be broadly explained in terms of an orientation sequence (see text).

Either explanation is plausible, but we prefer the latter, since an almost inevitable consequence of the geometry of the optically thick structure required by the unification scheme is that the nucleus will be partially obscured in some lines of sight. With this interpretation, the polar-scattered Seyfert 1 galaxies can be regarded as representing a transition between unobscured (the majority of type 1) and obscured (type 2) Seyferts. We can regard them as cases in which the inclination of the system axis to our line-of-sight is comparable to the half-opening angle of the torus.

### 5.1 Orientation-dependent polarization

Clearly, the combination of a compact (equatorial) scattering region within the torus and an extended (polar) scattering region outside it, will result in strongly orientation-dependent polarization properties. As outlined in S02, we can identify four inclination regimes, which produce quite distinct polarization signatures (Fig. 10). Although it is possible to use the GSM to predict precise inclination ranges for each of these regimes, the results would be sensitive to the poorly constrained parameters that define the two scattering regions. Therefore, we give approximate inclinations, assuming only that the torus half-opening angle is ~45°, as suggested by statistical arguments based on the relative numbers of type 1 and type 2 Seyferts (e.g. Schmitt et al. 2001b). We discuss below how the objects in our sample might be classified according to this scheme.

#### 5.1.1 *Weakly polarized Seyfert 1 galaxies ($i \approx 0°$)*

When $i \approx 0°$ (system viewed almost face-on), both the equatorial and polar scattering regions exhibit a high degree of circular symmetry in projection and therefore cancellation leads to null or very low polarization.

In S02, we identified 20 objects that exhibit clear evidence of polarization intrinsic to the active nucleus. Of the remaining 16 objects, most have average polarizations in the range $p \sim 0.5$–1 per cent, but have featureless polarization spectra and relatively high line-of-sight reddening and so could be seriously contaminated by interstellar polarization. However, there are 7 objects (Mrk 896, Mrk 926, NGC 3516, NGC 7213, NGC 7469, NGC 7603 and PG 1211+143) which have very low measured polarizations and must, therefore, be intrinsically weakly polarized regardless of any interstellar contamination. All have average polarizations ≤0.3 per cent which, given that the uncertainty is typically 0.1–0.3 per cent for our data, is broadly consistent with null polarization. Some of these objects could therefore be identified with the $i \approx 0°$ regime Note, however, that M96 measured a significantly higher level of polarization in NGC 3516, while NGC 7469 shows some evidence for a temporal variation in $p$ (S02). It should also be noted that other effects might also result in a low degree of polarization. It is possible, for example, that scattering column densities in both the equatorial and polar regions differ widely from object to object, leading to a wide variation in $p$ at any given inclination, or (less likely) that the equatorial and polar-scattered components cancel each other almost completely. Clearly, an independent constraint on orientation is needed to discriminate between such effects and pole-on orientation as the cause of their low polarizations. Moreover, the distinction between this category and the next is somewhat arbitrary, since in practice it is only possible to define 'null' polarization in terms of the signal-to-noise ratio of the data.

### 5.1.2 *Seyfert 1 galaxies dominated by equatorial scattering (0 < i < 45°)*

For inclinations in the range $0 < i < 45°$, there is no extinction along the direct line-of-sight to the nucleus. Both scattering regions, as well as the emission disc, are visible. The two scattering regions produce orthogonal polarization, so there will be cancellation between the two, and the net polarization will reflect that of the stronger component. From the relative frequencies with which Seyfert 1 galaxies exhibit polarization spectra characteristic of, respectively, equatorial- and polar scattering (Section 5.2), we infer that the equatorial component usually dominates.

The predicted polarization signatures of equatorial scattering of line emission from a rotating disc are a blue–red swing in $\theta$ across the H$\alpha$ profile and a central dip in polarisation in the line core, flanked by polarization peaks in the wings (Smith 2002; S04). In addition, the continuum and average H$\alpha$ polarization PA should be parallel to the projected radio axis. At least ten of the intrinsically polarized objects in the S02 sample show these features to differing degrees. Another object, not part of the S02 sample, which shows features consistent with equatorial scattering is NGC 4151 (M96).

The best example in this category is Mrk 6, which clearly exhibits all of the predicted polarization signatures. Others might exhibit, for example, a clear PA rotation across the H$\alpha$ profile, but not the predicted form of the percentage polarization variation, or vice versa (e.g. Mrk 985 and Mrk 509). However, as discussed in Smith (2002) and S04, the amplitudes of these effects depend on different combinations of parameters and the presence of one does not necessarily imply the presence of the other. There are also many complicating factors, such as light from the orthogonally-polarized polar-scattered component, which can distort the observed polarization signature. Perhaps the most curious case is Mrk 304 (S02), which exhibits the peak-trough-peak variation in $p$ over broad H$\alpha$, characteristic of equatorial scattering, but has a polarization PA perpendicular to its reported radio axis, characteristic of polar scattering. It is not clear if this is a polar scattering dominated object in which the scatterers have a large velocity dispersion, or an equatorial scattering dominated object in which part of the scattering disc is obscured. Either scenario could, in principle, explain the apparently conflicting polarization properties but, in our model, both require that the line-of-sight to the nucleus is at grazing incidence to the throat of the torus.

### 5.1.3 *Seyfert 1 galaxies dominated by polar scattering (i ≈45°)*

When $i \approx 45°$, the line-of-sight to the nucleus passes through the relatively tenuous upper layers of the torus, the equatorially-scattered light is attenuated and polar scattering dominates the polarization spectrum. The total light spectrum is still that of a Seyfert 1, albeit moderately reddened.

The prototype in this category is Fairall 51. Only 3 of the 20 intrinsically polarized Seyferts 1 galaxies in S02 are polar-scattered, but we have since identified several more, bringing the total to 12 (Section 2). In all but one of these objects the degree of polarization increases strongly towards the blue end of the spectrum with local maxima associated with the broad Balmer lines. The only exception is Was 45, in which the local peaks corresponding to the Hα and Hβ lines are present, but there is only a small increase in $p(\lambda)$ towards the blue. In all cases, $\theta(\lambda)$ shows little variation over the spectrum and tends to be roughly perpendicular to the PA of the radio source or ionisation cone, where these can be determined. The origin of the small PA rotations across the broad Hα line seen in some of the polar-scattered objects is unclear but one possibility is that after suffering extinction in the torus, there remains enough polarized flux from the equatorial scattering region to produce a small PA change when added to the dominant polar-scattered component.

Far-field polar scattering cannot by itself explain all the polarization properties of polar-scattered Seyfert 1 galaxies. As in PBL Seyfert 2 galaxies (Section 3), dilution of the polar-scattered light by unpolarized continuum emission is required to account for the observed wavelength dependence of $p(\lambda)$. In polar-scattered Seyfert 1 galaxies, the increase to the blue can be explained by dilution by reddened direct light (continuum and broad-line emission) from the active nucleus. However, to produce the local peaks in $p$ associated with the broad emission lines, dilution by a source whose spectrum does not contain these features is also necessary. This is likely to be the stellar continuum of the host galaxy (see, for example, discussion of Mrk 704 in GM94), which in the polar-scattered objects, will be somewhat stronger relative to the nuclear spectrum than in most Seyfert 1 galaxies, due to the extinction required to suppress the equatorially-scattered component. As already noted, Was 45 differs from the other polar-scattered Seyfert 1 galaxies in that $p(\lambda)$ only increases weakly to the blue. Assuming that its intrinsic nuclear spectrum is typical of a Seyfert 1, this suggests that the spectra of either the scattered light, or the diluting components (or both) have different slopes compared to the other objects. For example, a flatter $p(\lambda)$ spectrum would result if the scattered flux is itself significantly reddened, or if the stellar continuum is unusually blue.

### 5.1.4 Seyfert 2 galaxies (i >45°)

When the inclination is such that the line-of-sight to the nucleus passes through the body of the torus, only the polar scattering region remains unobscured and a Seyfert 2 nucleus, exhibiting broad-lines in polarized light, will be observed. In our simple model, 'polarized broad-lines' due to scattering in the polar cone should be present in all Seyfert 2 galaxies. However, only approximately one third exhibit polarized broad-lines at current sensitivity limits (e.g. Heisler, Lumsden & Bailey 1997; Moran et al. 2000; Lumsden & Alexander 2001). Indeed, it is currently unclear if all Seyfert 2 galaxies can be accommodated within the unification scheme − i.e. are really obscured Seyfert 1 galaxies.

Heisler et al. showed that the detectability of polarized broad Hα in Seyfert 2 galaxies is related to the ratio of *IRAS* fluxes at 60 μm and 25 μm ($f_{60}/f_{25}$); polarized broad lines only occurring in objects with 'warm' *IRAS* colours ($f_{60}/f_{25} < 4.0$). They interpret this as an orientation effect, in which the Seyfert 2 galaxies with 'cool' *IRAS* colours, are those seen at the most 'edge-on' orientations such that the torus itself obscures much of the polar scattering material, which is located, along with the hottest dust, within the inner throat of the torus. Conversely, Seyfert 2 galaxies with warm *IRAS* colours and polarized broad-lines have principal axes orientated closer to the line-of-sight, allowing a relatively clear view of the hot dust and the bulk of the polar scattering material. This scenario has been challenged on the grounds that absorbing column densities derived from X-ray data are similar for both warm and cool Seyfert 2 galaxies (Alexander 2001 and references therein), whereas in the Heisler et al. picture, cool objects should have higher column densities since the line-of-sight passes through a greater HI column (although see Antonucci 2002). Alexander (2001) suggests that given the large *IRAS* apertures, the $f_{60}/f_{25}$ ratio is more likely to be related to the presence or absence of current star formation in the environs of the active nucleus. In this case, the apparent absence of polarized broad-lines in many Seyfert 2 galaxies may be due to environmental effects such as dilution by a large starlight contribution or contamination by interstellar polarization. It would appear that more sensitive observations are required to establish the true prevalence of polarized broad-lines in Seyfert 2 nuclei and hence, whether or not all Seyferts can be incorporated in a generic unification scheme.

*5.1.5   Other objects*

There are a number of intrinsically polarized Seyfert 1 galaxies in our S02 sample whose status within the orientation sequence remains unclear. For example, several objects show a dip in polarization associated with the core of the broad Hα line but no evidence for any of the other features associated with equatorial scattering (e.g. ESO 141−G35, Mrk 279 and Mrk 290). Mrk 279 is particularly interesting since it exhibits a large (∼61°) change in *θ* over broad Hα. In addition, MS 1849.2−7832 shows some evidence for a PA rotation over the line. Better data might, perhaps, reveal clearer signatures of equatorial scattering in some of these objects.

Other objects exhibit significant variations in *p* and *θ* over broad Hα that are difficult to understand in terms of the simple orientation sequence illustrated in Fig. 10. For example, Mrk 335 exhibits a large PA rotation (∼80°), consistent with near-field equatorial scattering, but rather than the predicted peak-trough-peak variation, there is a general increase in *p* over the broad-line profile, with a prominent peak in the blue wing that appears to be associated with the PA rotation. Akn 564 has an optical polarization PA orthogonal to its projected radio source axes, as expected for polar scattering, but exhibits a dip in polarization over the broad Hα line. In other objects in which polar scattering is thought to dominate (Seyfert 2 galaxies and polar-scattered Seyfert 1 galaxies), *p* *increases* over the broad-line. The average polarization PA of the broad Hα line in NGC 5548 is orthogonal to the radio axis but also exhibits unique variations in *θ*, in that it appears to show two distinct rotations across the line profile. These objects may, perhaps, be explained by some combination of equatorial and polar scattering. However, it is also possible that isolated scattering centres (i.e. compact clouds of scattering material not associated with the polar or equatorial scattering regions) can produce enough polarized flux to significantly modify the net polarization spectrum. It would be unrealistic to expect the detailed distribution of scattering material to be the same in each Seyfert nucleus. Understanding such inter-object diversity would require detailed modelling of individual sources and the results would not necessarily be generally applicable. Our aim here is to develop a general framework, which can explain the broad range of polarization spectra exhibited by Seyfert nuclei, rather than the detailed properties of individual objects.

## 5.2   Relative frequency of polar-scattered Seyfert 1 galaxies

Having established that both polar and equatorial scattering routes are present in Seyfert 1 galaxies, and hence by implication of the unification scheme, all Seyfert galaxies, it is of interest to gain an insight into the relative numbers of null-polarization, equatorially-scattered and polar-scattered objects. In the context of the two-component scattering geometry outlined above, this information can be used to infer inclination ranges for these polarization/orientation classes just as the relative frequency of Seyfert 1 galaxies and Seyfert 2 galaxies can be used to deduce an average torus opening angle (e.g. Lawrence 1991). Since our model implies that Seyfert 1 polarization properties are orientation-dependent and indeed, can be modified by dust extinction, any such analysis is likely to be subject to selection effects similar to those that have plagued comparative studies of Seyfert 1 and 2 galaxies (e.g. Schmitt et al. 2001b). The largest and arguably least biased spectropolarimetric survey of Seyfert 1 galaxies is that of S02, which thus provides the most suitable existing sample for assessing the relative frequencies of the different polarization classes. Even so, a number of caveats should be borne in mind when considering the results of such an analysis. Firstly, it is magnitude limited in *V* and, since we have estimated that the polar-scattered Seyfert 1 galaxies suffer 1–4 magnitudes of visual extinction through the atmosphere of the torus, it is possible that these objects are under-represented. Secondly, and probably more seriously, equatorially-scattered objects tend to have lower overall polarizations than polar-scattered objects and therefore a higher signal-to-noise ratio is required to identify them. For the same reason, equatorially-scattered objects are likely to be more easily masked by foreground interstellar polarization. Conceivably, a number of equatorially-scattered objects might have been missed among the 16 Seyfert 1 galaxies in which we were unable to detect evidence of polarization intrinsic to the active nucleus. The same considerations are, of course, even more pertinent for null-polarization objects.

Nevertheless, while recognizing that these effects may distort the results, we have used our S02 sample to estimate the relative frequencies of the three Seyfert 1 polarization classes we have identified. We include in this analysis the 20 objects identified as being intrinsically polarized (i.e. those whose nuclear polarization is clearly not dominated by foreground interstellar polarization arising in either our own Galaxy or the host) and also the 7 'null-polarization' objects that have measured polarizations, averaged over wavelength, $p$ <0.3 per cent (see Section 5.1.1). The division among the three Seyfert 1 polarization/orientation classes is as follows. Out of the 27 objects, 7 have null polarization, 10 are equatorially-scattered, 3 are polar-scattered and the remaining 7, while intrinsically polarized, have polarization spectra whose classification within our model remains unclear. As we have already noted, the distinction between null-polarization and equatorially-scattered objects is a function of signal-to-noise. This is probably also true for some of the 'unclassified' objects. In comparison, the polar-scattered Seyfert 1 galaxies are much more homogeneous in their polarization properties and (usually) more highly polarized. They are, therefore, the easiest group to find and identify. It follows that the most robust statistic that we can extract from this analysis is the fraction of polar-scattered objects amongst Seyfert 1. Our sample suggests that this is ~10 per cent. Allowing that 1 in 3 Seyferts is a type 1 (Schmitt et al. 2001b), polar-scattered Seyfert 1 galaxies make up only ~3 per cent of the population. In terms of the torus structure, this implies that ~3 per cent of the sky, as seen from the central source, is covered by the semi-opaque atmosphere. This in turn implies that the arc of inclination, within which polar scattering dominates the polarization is $\Delta i \sim 2°$ (assuming a torus opening angle of $45°$).

However, we have identified a further 9 polar-scattered Seyfert 1 galaxies in this paper and there are indications that these objects may be more common in other studies (although they had not previously been identified as a group with common polarization properties). It is, therefore, of interest to estimate the relative frequency of polar-scattered Seyfert 1 galaxies more generally. To do this, we have compiled an extended sample of Seyfert 1 galaxies from Goodrich (1989a,b); GM94; M96; Schmid et al. (2000, 2003); S02 and the new observations presented in this work. Whilst this is not a comprehensive list, it includes the major works in the field and we believe that between them, these papers contain nearly all the good quality optical spectropolarimetry of Seyfert 1 galaxies that has been published to date. Smith et al. (1997) obtained UV and optical spectropolarimetry of I Zw1 and Mrk 486. However, the polarization PA's taken from S02 and M96 for these objects are in good agreement with the measurements of Smith et al.

We have only included objects in which foreground interstellar polarization arising in either our own Galaxy or the host galaxy appears to be insignificant compared to the intrinsic polarization of the AGN, rather. For our own sample, and that of M96, this was assessed by visual inspection of the polarization spectra, using the same criteria as S02. However, in GM94 and Goodrich (1989a,b), $p$ and $\theta$ spectra are only presented for selected objects and for the remaining sources in these papers we have relied on the author's own assessment as to whether they are intrinsically polarized (the criterion usually being $p >$ 0.5 per cent).

Thus excluding objects that may be dominated by interstellar polarization and accounting for duplications, the extended sample numbers 42. This sample is listed in Table 4. Objects exhibiting signatures of polar or equatorial scattering are designated 'POL' or 'EQ', respectively, in column 3, while references to the polarization data are given in column 7. There are 19 objects to which we cannot confidently assign a polarization class. These include objects for which $p$ and $\theta$ spectra were not published, or whose spectra have a relatively poor signal-to-noise ratio, as well as a handful with good signal-to-noise data, but which do not exhibit characteristics predicted by our basic model. Note that, because of the inhomogeneity of the various data-sets from which the sample is drawn, we have not included any null-polarization candidates.

As discussed in Section 2, 12 objects (29 per cent) have polarization characteristics consistent with polar scattering. We have identified signatures of equatorial scattering in one object (NGC4151 − M96) in addition to those already identified in S02, making a total of 11 (26 per cent) equatorial scattering dominated objects. At first sight, therefore, it would seem that Seyfert 1 galaxies dominated by, respectively, polar scattering and equatorial scattering are found with approximately equal frequency, each category making up slightly more than 25 per cent of the sample. However, it is almost certain that polar scattering dominated Seyfert 1 galaxies are over-represented in the extended sample, while equatorially-scattered objects are under-represented. As already noted, the latter tend to be more weakly polarized and hence, are harder to detect at a given signal-to-noise. Moreover, several of the studies we have used are, by their nature, biased in favour of polar-scattered Seyfert 1 galaxies. For example, GM94 and M96 explicitly studied high-polarization Seyfert 1 galaxies, while the Seyfert 1.8 and 1.9 galaxies studied by Goodrich (1989a) would naturally fall into the polar-scattered category in our model (Section 5.4). Therefore, taking into account also the omission of weakly polarized objects, it is reasonable to regard the fraction of polar-scattered objects in the extended sample, 29 per cent, as a firm upper limit for their occurrence amongst Seyfert 1 galaxies. They thus make up less than about 10 per cent of the Seyfert population and, with the assumptions used above, this corresponds to a maximum inclination range $\Delta i \sim 8°$.

### 5.3 Polarization position angle alignments

A clear prediction of our model is that the polar and equatorial scattering dominated objects should be orthogonally polarized. Given that the equatorial scattering region is co-axial with the torus and that the illumination cone defined by the latter is uniformly filled by scattering particles, the polarization **E** vector will be perpendicular to the projection of the torus axis if polar scattering dominates, and parallel to it if equatorial scattering dominates. An observational test of this prediction requires an independent indicator of the orientation on the sky of the projected torus axis. It has traditionally been assumed that the radio source axis traces the symmetry axis of the AGN and several earlier studies have compared the optical polarization PA with that of the radio source in Seyferts (Antonucci 1983; 2001; Brindle et al. 1990; M96). These studies employed relatively small samples (sample size being limited by the availability of both polarization data and a well-determined radio axis), but they generally agree that the optical polarization in Seyfert 1 galaxies tends to be parallel to the radio source, though there are also some objects in which it is orthogonal. Our model provides a simple explanation for the mixture of 'parallel' and 'perpendicular' polarizations found amongst Seyfert 1 galaxies in terms of the dominant scattering route.

With our new data, we can attempt to quantify the fractions of Seyfert 1 galaxies in the 'parallel' and 'perpendicular' groups in a larger sample. We have searched the literature for information on the radio source morphology of the objects included in the extended Seyfert 1 spectropolarimetry sample listed in Table 4. Unfortunately, it is often difficult to establish the radio axis unambiguously. In many Seyfert 1 galaxies the radio source is unresolved or marginally resolved in existing images. In others, the radio source PA on large (kiloparsec) scales differs from that on small (parsec) scales. We have used the small-scale axis where available as this presumably better represents the symmetry axis of the system, whereas environmental effects are more likely to influence the kiloparsec-scale morphology. We have taken the published PA where this is given explicitly. Otherwise, if a radio map is available, the PA was determined by eye to an accuracy of about ±10° in those objects in which the radio source extends along a preferred axis. In the absence of obvious jet-like features or double and triple source components the PA was determined from the innermost contour tracing an elongation in the source. By these methods, we have obtained radio source PA's for 22 out of the 42 objects in the extended sample. In Table 4 are listed the polarization PA[1] of the optical continuum (column 4), the radio source PA (column 5) and the difference between the two, $\Delta PA$ (column 6). However, as we have used radio maps from heterogeneous sources we have also compiled a subset including only the 12 objects in which the radio source axis can be determined unambiguously. These are objects, designated 'L' in column 8, exhibiting linear structures, namely a clear radio jet (aspect ratio >2:1) or well-resolved and aligned double/triple source components. The other designations used in column 8 to describe the radio source morphology are PC: position angle cited as private communication based on unpublished data; A: ambiguous structure; S: slightly resolved. References to radio data are given in column 7.

For the purpose of comparing the optical polarization PA with the radio source orientation, we divide the sample into three categories based on the magnitude of $\Delta PA$. *Parallel* objects have $\Delta PA \leq 30°$, *perpendicular* objects have $\Delta PA \geq 60°$ and *intermediate* objects have $30° < \Delta PA < 60°$. These bins in $\Delta PA$ are, of course, arbitrary, and adjusting their boundaries by only 1 or 2° would cause some objects to change category. This does not, however, significantly affect the overall conclusion. Of the 22 objects with known radio source PA's, 10 are parallel (45 per cent), 5 are perpendicular (23 per cent) and 7 intermediate (32 per cent). If we count only the subset with clear linear radio sources, we find rather similar results: 5 objects are parallel (42 per cent), 3 are perpendicular (25 per cent) and 4 are intermediate (33 per cent).

---

[1] The polarization PA was measured from the continuum adjacent to Hα, where possible (M96; S02; this work). Otherwise, the quoted value is used. In cases where multi-epoch observations exist, the PA measurement is taken from the data having the highest apparent signal-to-noise.

As we have already noted, the extended spectropolarimetry sample is probably strongly biased in favour of polar-scattered Seyfert 1 galaxies and hence objects with perpendicular polarization may be over-represented. Our original Seyfert 1 sample was selected on the basis of optical magnitude, without regard to prior knowledge of polarization properties (S02). Whilst this sample may also be biased in favour of polar-scattered objects (as noted in Section 5.2) it is likely to be less so than the extended sample. The S02 sample contains 20 intrinsically polarized objects of which 10 have known radio source PA's. Five of these are parallel objects, 2 are perpendicular and 3 are intermediate. Unfortunately, there are only 6 objects with jet-like radio sources: 2 parallel, 1 perpendicular and 3 intermediate.

**Table 4.** Polarization and radio position angles for the extended spectropolarimetric sample. The first reference given in column 7 is for the optical continuum polarization PA: (1) S02; (2) This work; (3) M96; (4) Goodrich (1989a); (5) Goodrich (1989b); (6) Schmid et al. (2000); (7) Schmid, Appenzeller & Burch (2003). The second reference is for the radio source PA: (8) Condon et al. 1998; (9) Moran (2000); (10) Kellerman et al. (1994); (11) Blundell & Lacy (1995); (12) Capetti et al. (1995b); (13) Thean et al (2001); (14) M96; (15) Singh and Westergaard (1992); (16) Wilson and Ulvestad (1982); (17) Ulvestad et al. (1999); (18) Mundell et al. (1995); (19) Thean et al. (2000); (20) Mijayi et al. (1992); (21) Christopoulou et al. (1997); (22) Mundell et al. (2003); (23) Nagar et al. (1999); (24) Antonucci (1985); (25) Ulvestad & Wilson (1984a).

| Object | IAU Name | Pol. Class | Pol. PA (°) | Radio PA (°) | ΔPA (°) | Refs | Rad Morph. |
|---|---|---|---|---|---|---|---|
| Akn 120 | J0516−00 | EQ | 77.0±1.1 | 50 | 24.0 | 1,8 | S |
| Akn 564 | J2242+29 |  | 87.0±1.3 | 170 | 83.0 | 1,9 | L |
| ESO 141−G35 | J1913−60 |  | 179.1±0.6 |  |  | 1 |  |
| Fairall 51 | J1844−62 | POL | 141.2±0.2 |  |  | 1 |  |
| IZw1 | J0053+12 | EQ | 151.6±0.5 | 140 | 8.9 | 1,10 | S |
| KUV 18217+6419 | J1821+64 | EQ | 143.3±2.0 | 20 | 56.7 | 1,11 | L |
| Mrk 6 | J0652+74 | EQ | 156.5±0.8 | 170 | 13.5 | 1,12 | L |
| Mrk 279 | J1353+69 |  | 58.9±2.4 | 96 | 37.1 | 1,13 | L |
| Mrk 290 | J1535+57 |  | 157.5±1.1 |  |  | 1 |  |
| Mrk 304 | J2217+14 | EQ | 135.5±2.5 | 42 | 86.5 | 1,14 | PC |
| Mrk 335 | J0006+20 |  | 113.6±1.7 |  |  | 1 |  |
| Mrk 509 | J2044−10 | EQ | 151.8±0.9 | 165 | 20.9 | 1,15 | L |
| Mrk 841 | J1504+10 | EQ | 103.4±1.0 |  |  | 1 |  |
| Mrk 876 | J1613+65 | EQ | 110.5±1.4 | 140 | 29.5 | 1,10 | S |
| Mrk 985 | J0120+38 | EQ | 114.3±0.4 |  |  | 1 |  |
| MS 1849.2−7832 | J1857−78 |  | 7.0±1.1 |  |  | 1 |  |
| NGC 3783 | J1139−37 | EQ | 135.5±1.0 |  |  | 1 |  |
| NGC 4593 | J1239−05 | POL | 149.9±5.1 |  |  | 1 |  |
| NGC 5548 | J1417+25 |  | 33.2±0.5 | 165 | 48.2 | 1,16 | L |
| Was 45 | J1204+31 | POL |  |  |  | 1 |  |
| Mrk 231 | J1256+56 | POL | 95±0.4 | 5 | 90 | 2,17 | L |
| NGC 3227 | J1023+19 | POL | 122.4±1.0 | 170 | 47.6 | 2,18 | L |
| Mrk 9 | J0736+58 |  | 138 | 105 | 33 | 3,19 | A |
| Mrk 352 | J0059+31 |  | 135 |  |  | 3 |  |
| Mrk 376 | J0714+45 | POL | 173 |  |  | 3 |  |
| Mrk 486 | J1536+54 |  | 130 |  |  | 3 |  |
| Mrk 704 | J0918+16 | POL | 64.5 |  |  | 3 |  |
| Mrk 1048 | J0234−08 |  | 106.5 | 105 | 1.5 | 3,8 | S |
| NGC 3516 | J1106+72 |  | 173 | 10 | 17 | 3,20 | L |
| NGC 4051 | J1203+44 |  | 85 | 73 | 12 | 3,21 | L |
| NGC 4151 | J1210+39 | EQ | 91 | 77 | 14 | 3,22 | L |
| IRAS 19580−1818 | J2000−18 |  | 155.2±0.4 |  |  | 4 |  |
| Mrk 883 | J1629+24 |  | 153±6 | 110 | 43 | 4,8 | S |
| Mrk 1218 | J0838+24 | POL | 65.7±0.3 |  |  | 4 |  |
| Mrk 507 | J1748+68 |  | 11.6±1.3 |  |  | 5 |  |
| Mrk 766 | J1218+29 | POL | 90.0±0.3 | 27 | 63 | 5,23 | S |
| Mrk 957 | J0041+40 |  | 42.7±2.8 | 50 | 7.3 | 5,24 | S |
| Mrk 1126 | J2300−12 |  | 172.7±2.7 | 100 | 72.7 | 5,25 | L |
| Mrk 1239 | J0952−01 | POL | 130.1±0.2 |  |  | 5 |  |
| IRAS 15091−2107 | J1511−21 | POL | 61.7±0.2 | 2 | 59.7 | 5,19 | S |
| ESO 198−G24 | J0238−52 |  | 166 |  |  | 6 |  |
| ESO 323−G077 | J1306−40 | POL | 84.7 |  |  | 7 |  |

Parallel orientations, therefore, outnumber the perpendicular cases by roughly 2:1 in both the extended and S02 samples, regardless of whether we consider all available radio PA's or only the jet-like radio sources. However, the distribution in *ΔPA* is not bi-modal; intermediate objects are as common as perpedicular ones and overall, parallel orientations account for only 50 per cent of the sample. This is true even if only jet-like radio sources are considered, suggesting that the intermediate cases are not simply the result of a mis-identification of the radio axis. In some cases, contamination by foreground interstellar polarization may significantly affect the measured PA. Another possibility is that the radio jet is itself misaligned with the optical symmetry axis as defined by either the torus or the emitting/scattering disc system. Alternatively, as discussed in more detail below, the polarization PA can be offset from that of the radio source if the polar scattering region has an asymmetric density distribution.

The fraction of perpendicular objects (~25 per cent) is broadly consistent with the fraction of polar-scattered Seyfert 1 galaxies found in the extended spectropolarimetry sample. However, we would also expect a detailed correspondence between the presence of polar scattering signatures in the polarization spectrum and a perpendicular relationship between the polarization PA and the radio axis. There should be a similar correspondence between signatures of equatorial scattering and parallel alignments. As we have already discussed (Section 2.3), the polar-scattered Seyfert 1 galaxies generally fall into the perpendicular group when the radio axis PA can be estimated. The only notable exception is NGC 3227, in which the polarization PA is offset by 45° relative to the radio axis, but is approximately orthogonal to the major axis of the extended narrow line region. In order to explain the misalignment between the radio source and the [OIII] emission, Mundell et al. (1995) propose that the latter traces the intersection, with the disc of the galaxy, of the conical UV radiation beam escaping from the torus. Therefore, although the torus and radio source share a common axis, the radiation cone is only partially filled with gas and thus, viewed in projection, the line-emitting gas and the radio source are not aligned. We would expect the scattering material (electrons or dust) to be closely associated with the line-emitting gas. In this case, the scattering plane will be aligned with the latter, rather than the radio axis, giving a polarization PA perpendicular to the extended emission line region, but misaligned with the projected axis of the radio source. Therefore, although not perpendicular to the radio axis, the polarization PA in NGC 3227 is, never the less, consistent with polar scattering.

The example of NGC 3227 shows that a significant misalignment between the polarization PA and the perpendicular to the radio axis can occur as a result of an asymmetric density distribution within the polar scattering cone. It is possible that the intermediate polarization PA's in other objects have a similar explanation. However, NGC 3227 itself is the only one in this category that exhibits the spectral signatures of polar scattering.

There are a number of objects with 'perpendicular' polarization PA's that are not among the polar-scattered Seyfert 1 galaxies discussed in Section 2. However, none of these provide compelling counter examples. One, Mrk 304, actually exhibits spectropolarimetric signatures of equatorial scattering (see Section 5.1.2), but the radio PA in this case is taken second-hand from M96, who cites a private communication based on unpublished observations, so the reliability of the quoted value is unclear. Conversely, both Akn 564 and Mrk 1126 have reasonably secure radio axis PA's, but uncertain spectropolarimetry. The former shows an apparent dip in *p* associated with the blue side of the broad Hα line (S02), but the signal-to-noise ratio in these data is relatively low. The wavelength coverage is also insufficient to determine if *p* increases to the blue in the continuum. Goodrich (1989b) remarks that the continuum polarization in Mrk 1126 increases to the blue, but the polarization spectra are not shown.

We have been able to obtain radio source PA's for 8 of the 11 equatorial scattering dominated objects. All but two have parallel polarization

PA's, as predicted. The exceptions are Mrk 304, which we have already discussed, and KUV 18217+6419. The latter has a relatively low level of polarization (~0.3 per cent) and could therefore be affected by interstellar polarization. However, it also exhibits a very large PA swing (~70°) across the broad H$\alpha$ line, such that the blue and red wing polarizations are approximately perpendicular to and parallel to the radio source axis, respectively (S02). This object clearly has complex polarization properties that will repay further study.

There are 4 objects with parallel polarization PA's that (on the basis of existing data) we have not categorized as equatorial scattering dominated. The radio sources in Mrk 1048 and Mrk 957 are only marginally resolved. In the former, spectropolarimetry obtained by M96 shows variations in $p$ and $\theta$ across broad H$\alpha$, suggesting that this object may indeed be equatorially-scattered. In the case of Mrk 957, the polarization PA is taken from Goodrich (1989b) who, however, does not present spectra so we are unable to ascertain the nature of its broad H$\alpha$ polarization. The two remaining objects, NGC 4051 and NGC 3516, both have well-determined radio PA's. There is no firm evidence for polarization structure in the broad H$\alpha$ line in NGC 4051, but the data are relatively noisy (M96; S02). In the case of NGC 3516, our results differ significantly from those of M96. We find this object to be essentially unpolarized, with an average $p=0.1\pm0.04$ per cent (S02). M96, in contrast, measured $p\sim1$ per cent and found variations in both $p$ and $\theta$ over broad H$\alpha$, which *are* consistent with equatorial scattering. The reason for the disagreement between the two data sets is unclear.

### 5.4 Polar scattering among AGN types

A number of different sub-types of Seyfert nuclei have been identified according to their spectroscopic characteristics. Apart from the distinction between Seyfert 1 and 2 galaxies, based on whether or not broad-lines can be discerned in the total flux spectrum, Seyfert 1 galaxies are themselves sub-classified as types 1.5, 1.8 and 1.9 according to the (decreasing) prominence of the Balmer-line wings relative to the narrow lines (Osterbrock 1981). In addition, the so-called Narrow-line Seyfert 1 galaxies (NLS1; Osterbrock & Pogge 1985), are characterized by extreme optical and X-ray properties, manifested most obviously in the optical spectrum by strong but relatively narrow (FWHM <2000 kms$^{-1}$) broad Balmer lines. All of these spectroscopic sub-types are represented in one or more of the three Seyfert 1 polarization/orientation classes discussed above.

In S02 we argued that the H$\alpha$ polarization properties of the NLS1 in our sample are indistinguishable from those of the general population of Seyfert 1 galaxies. There are 8 NLS1 in the S02 sample, three of which we deem to be intrinsically polarized. None of these shows clear signatures of polar scattering. In contrast, of the 17 NLS1 studied by Goodrich (1989b), only 6 are considered to be intrinsically polarized, but at least 3 of these, Mrk 766, Mrk 1239 and *IRAS* 1509–211 are polar-scattered according to our criteria (Section 2). In addition, Goodrich notes that the continuum polarization increases to the blue in two more objects (Mrk 507 and Mrk 1126), for which polarization spectra were not presented. Thus, possibly as many as 5 of the 6 intrinsically polarized NLS1 in Goodrich's sample are polar-scattered. Combining the two samples yields a total of 24 NLS1 (there is one object in common), of which 9 are intrinsically polarized. Therefore, between 12 and 21 per cent of NLS1 are dominated by polar scattering, consistent with the limits we deduced from the S02 and extended samples for Seyfert 1 galaxies in general. In contrast, only one NLS1, IZw1 shows signatures of equatorial scattering, suggesting that objects dominated by this scattering route might be under abundant amongst NLS1. However, while both samples should be relatively unbiased with respect to polarization properties, as we have already noted, the lower levels of polarization typically exhibited by equatorially-scattered objects makes them harder to detect than polar-scattered objects at a given signal-to-noise ratio. It is possible, therefore, that with better data several more equatorially-scattered objects will be revealed amongst the low polarization NLS1. It is clear, never the less, that polar-scattered objects are at least as common amongst NLS1 as they are amongst Seyfert 1 in general. In the context of the two-component scattering geometry, polar-scattered objects are viewed at a large angle ($i\sim45$) to the symmetry axis. This argues against models in which the relatively small broad-line widths of NLS1 are due to a preferred face-on orientation (Puchnarewicz et al. 1992) and favours those in which the broad-lines are intrinsically narrow (e.g. Pounds et al. 1995).

Seyfert 1.8 and 1.9 galaxies have relatively steep broad-line Balmer decrements, suggesting that the BLR is highly reddened and hence that dust extinction along the line-of-sight to the AGN is responsible for the weakness of the broad Balmer-line wings (Goodrich 1990). The range in visual extinction implied by the observed Balmer decrements is $A_V$~1–4. This is comparable with the range over which we would expect polar scattering to dominate the polarization in Seyfert 1 galaxies, before the broad H$\alpha$ line in direct light is suppressed (Section 4.2). Moreover, the large variations in the broad-line fluxes over timescales of several years that have been observed in some objects are consistent with changes in reddening (Goodrich 1989a). In some cases, the variation in broad-line flux has been large enough to change the spectroscopic classification of the nucleus from 1.8 or 1.9 to 1, as in the case of Mrk 1218 (Goodrich 1989a and references therein). A plausible explanation for these properties is that Seyfert 1.8 and 1.9 nuclei are viewed through the patchy and turbulent atmosphere of the circum-nuclear torus and, therefore, they naturally fit into our two-component scattering model as polar scattering dominated Seyfert 1 galaxies in which the optical depth along the line-of-sight to the central structures (i.e. the continuum source, BLR and equatorial scattering region) approaches the value needed to suppress the direct broad-line flux. The higher optical depth (as compared to polar-scattered Seyfert 1 galaxies with prominent broad-lines) could result either from a higher inclination (and hence a longer path length through the torus atmosphere) or from a transient increase due to the passage of a particularly optically thick cloud across the line-of-sight.

It follows that Seyfert 1.8 and 1.9 galaxies should exclusively exhibit spectropolarimetric signatures of polar scattering. Only one Seyfert 1.8, Was 45, is included in the S02 sample and as discussed in Section 2, we believe its polarization spectrum to be consistent with polar scattering. Goodrich (1989a) studied the optical polarization of 16 Seyfert 1.8 and Seyfert 1.9 galaxies but found only three (Mrk 1218, *IRAS* 19580–1818 and Mrk 883) that are clearly intrinsically polarized. Mrk 1218 displays clear polar scattering characteristics (Section 2), although it was actually in a Seyfert 1 state when the spectropolarimetry was obtained. The behaviour of $p(\lambda)$ and $\theta(\lambda)$ in Mrk 883 cannot be determined since Goodrich's (1989a) spectropolarimetry yielded only wavelength-averaged values. *IRAS* 19580–1818 has an average continuum polarization of about 3 per cent and is described as having a highly polarized broad H$\alpha$ line. These properties are consistent with it being dominated by polar scattering, but since $p$ and $\theta$ spectra are not presented and there is no information on the radio axis PA, we cannot verify this. It is perhaps surprising, in the context of this interpretation, that only 3 out of 16 objects were found to be intrinsically polarized, given that polar-scattered Seyfert 1 galaxies are typically polarized at levels ~1 per cent or higher. However, we note also that only 1/3 of Seyfert 2 galaxies have been found to exhibit polarized broad-lines, despite much more intensive searches. The explanations posited for the 'missing' PBL Seyfert 2 galaxies (Alexander 2001) may well also apply to Seyfert 1.8 and 1.9 galaxies.

Seyfert 1.8 and 1.9 galaxies are sometimes included in searches for polarized broad-lines in Seyfert 2 nuclei (e.g. Lumsden et al. 2001). Indeed, the extent to which the faint broad-lines seen in the total flux spectra are due to scattering is unclear – in some cases, faint broad wings observed in total flux may be entirely scattered light, as in the radio galaxy 3C234 (Young et al. 1998). This makes the distinction between Seyfert 1.9 (in particular) and Seyfert 2 somewhat hazy. In any case, our work points to a smooth progression between Seyfert 1 and 2 characteristics governed by increasing inclination and hence, for $i>45°$; an increasing path length through the torus.

In comparison with Seyfert galaxies, relatively little is known in general about the broad-line polarization of their high luminosity counterparts, radio-quiet quasars (RQQ's). While the basic unification scheme is well established for Seyfert galaxies, it is less certain that it can be simply extended to higher luminosities. For example, while the long standing problem of the 'missing' obscured 'type 2' RQQ's has been at least partly resolved by the discovery of plausible candidates among the Ultra-Luminous Infra-red Galaxies (ULIRG's; e.g. Hines et al. 1999b; Antonucci 2001) and in recent X-ray surveys (e.g. Norman et al. 2002; Stern et al. 2002), it is not yet clear that these objects exist in the numbers predicted by a simple extrapolation of the Seyfert unification scheme. The structure and geometrical parameters of the torus or even its existence in the form envisaged in the unification scheme may be functions of luminosity. So, similarly, might the spatial distribution and density of the scattering medium. For this reason, it is of interest to ask if our two component scattering model can be applied to RQQ's and if so, to determine the distribution of objects amongst the various polarization/orientation classes as a function of luminosity. Five of the objects in our S02 sample are luminous enough to be classified as quasars according to the arbitrary, but often used, criterion $M_B < -23$. None of these are polar scattering dominated. In fact, 3 exhibit signatures of equatorial scattering (KUV 18217+6419, Mrk 509 and Mrk 876) and the remaining 2 (Mrk 926 and PG1211+143) are intrinsically weakly polarized. The polarization properties of these objects are thus consistent with them being viewed at relatively low inclinations ($i < 45$).

However, recent spectropolarimetric observations of a sample of reddened, high-luminosity AGN detected in the Two-Micron All Sky Survey (2MASS) have revealed a range in both spectroscopic and spectropolarimetric properties broadly comparable to what is seen in optically-selected Seyfert 1 galaxies (Smith et al. 2003). This sample may well furnish several candidates for high-luminosity, polar scattering dominated broad-line AGN; there is at least one type 2 object that exhibits prominent broad lines in polarized flux.

Our two-component scattering model and the corresponding polarization/orientation sequence for Seyferts outlined in Section 5.1, bear certain similarities to models that aim to account for the spectral and polarization properties of Broad Absorption Line QSO's (BALQSO's; e.g. Glenn, Schmidt & Foltz 1994; Hines & Wills 1995; Goodrich & Miller 1995; Cohen et al. 1995; Schmidt, Hines & Smith 1997). These postulate that BALQSO's are a subset of RQQ's viewed through the 'skin' of the obscuring torus, with the clouds responsible for the broad absorption line features being ablated from the surface of this structure (e.g. Weymann et al. 1991). Wills & Hines (1997) and Schmidt & Hines (1999) further propose that non-BALQSO's, BALQSO's and narrow-lined Hyper-luminous ULIRG's (HIG's) form a '3-bin' orientation sequence with BALQSO's representing the transition between unobscured (non-BALQSO's) and highly obscured objects (HIG's). In this context, BALQSO's could be regarded as high luminosity counterparts of polar-scattered Seyfert 1 galaxies. Interestingly, both Mrk 231 and Mrk 1239 show evidence for scattering outflows (Section 2.3) and Mrk 231 also exhibits a low-ionization, blue-shifted absorption line system (Smith et al. 1995) reminiscent of a low ionization BALQSO. However, a comparison of polarization properties suggests that polar-scattered Seyfert 1 galaxies are not simply low luminosity BALQSO's. While the optical continuum polarization in BALQSO's exhibits characteristics similar to those seen in polar-scattered Seyfert 1 galaxies (BALQSO's tend to be more highly polarized than non-BALQSO's, with the polarization increasing to the blue) the broad emission lines show rather different polarization properties (e.g. Ogle et al. 1999). In BALQSO's, the broad emission lines are often polarized at a lower level than the continuum and at a different PA, suggesting a more complex scattering geometry than appears to be the case in polar-scattered Seyfert 1 galaxies.

At the lower end of the AGN luminosity scale, our scattering model may also have an application to Low-Ionization Emission Line Region Galaxies (LINERs). Barth et al. (1999) have shown that several LINER nuclei exhibit polarized broad-lines and propose that these lower luminosity objects may have a circum-nuclear geometry similar to that envisaged in the Seyfert unification scheme. Of the three objects in which polarized broad-lines are detected, two (NGC 315 and NGC 1052) exhibit faint broad H$\alpha$ lines in total flux. Both of these objects exhibit increases in $p$ across broad H$\alpha$ and have optical polarization PA's approximately orthogonal to their radio axes, both properties that are consistent with polar scattering.

## 6 CONCLUSIONS

We have previously reported the results of an extensive study of the optical polarization spectra of Seyfert 1 nuclei (S02). The intrinsically polarized objects in this sample commonly show variations in $p$ and/or $\theta$ over the broad H$\alpha$ line that can be attributed to scattering of light from a rotating emission disc by a surrounding, compact scattering disc in the equatorial plane of the circum-nuclear torus. Here, we identify and discuss a significant minority of Seyfert 1 galaxies that appear to be dominated by scattering in an extended region along the poles of the torus. These objects exhibit the following characteristics, which distinguish them from the majority of Seyfert 1 galaxies:

i.   A systematic increase in $p(\lambda)$ towards shorter wavelengths, with local peaks associated with the broad Balmer lines;
ii.  $\theta(\lambda)$ does not exhibit large changes across the broad emission lines
iii. The average polarization position angle is perpendicular to the projected radio source axis.

These characteristics are also typical of Seyfert 2 nuclei in which polarized broad-lines have been detected. The 'polar-scattered' Seyfert 1 galaxies therefore provide direct evidence that scattering in the 'illumination cone' of the torus, the mechanism responsible for polarized broad-lines in Seyfert 2 galaxies, also occurs in Seyfert 1 galaxies. We infer from this that, broadly speaking, all Seyferts have similar scattering geometries, as would be expected in the Seyfert unification scheme.

We propose that all Seyfert nuclei have both equatorial and polar scattering regions located, respectively, inside and outside the torus, and producing orthogonal polarization. For both components, the degree of polarization increases monotonically with inclination, implying that the dominant state is determined by factors other than the average scattering angle. In most Seyfert 1 galaxies, the observed polarization appears to be dominated by equatorial scattering. We argue that the polar-scattered Seyfert 1 galaxies are oriented such that our line-of-sight to the nucleus passes through the upper layers of the torus and is subject to a moderate amount of extinction; enough to suppress polarized light from the equatorial scattering region, though not the broad wings of the Balmer lines. The polar-scattered Seyfert 1 galaxies, therefore, represent a transition state between unobscured (the majority of type 1) and obscured (type 2) Seyferts.

The range of polarization properties exhibited by Seyfert galaxies can be broadly understood in terms of an orientation sequence based on the two-component scattering model.

i. When the system is viewed almost face-on ($i \approx 0°$), both the equatorial and polar scattering regions exhibit a high degree of circular symmetry and cancellation leads to null or intrinsically weak polarization.
ii. At intermediate inclinations ($0 < i < 45°$), there is no extinction along the direct line-of-sight to the nucleus and both scattering regions, as well as the broad-line region, are visible. In general, equatorial scattering dominates the observed polarization.
iii. When the inclination of the system axis is comparable to the torus opening angle ($i \approx 45°$) the line-of-sight to the nucleus is subject to a moderate amount of extinction ($A_V \sim 1-4$ mag), and polar-scattered Seyfert 1 galaxies are observed.
iv. At still larger inclinations ($i > 45°$), both the BLR and equatorial scattering region are completely obscured by the torus and the broad-lines are only visible in polarized light scattered from the polar scattering region. A Seyfert 2 with polarized broad-lines is observed.

We estimate that between 10 and 30 per cent of Seyfert 1 galaxies are dominated by polar scattering, representing 3–10 per cent of all Seyferts. Examples of polar-scattered Seyfert 1 galaxies are found amongst all of the main spectroscopic sub-types (Seyfert 1, 1.5, 1.8 and 1.9), pointing to a smooth progression between Seyfert 1 and 2 characteristics governed by increasing inclination which, for $i > 45°$, gives an increasing path length through the torus. In the context of our model, the fraction of polar-scattered Seyfert 1 galaxies implies that the range in inclination, centred around $i \approx 45°$, over which partial obscuration of the AGN occurs is $\Delta i \approx 3-8°$.

Polar-scattered Seyfert 1 galaxies are as common amongst NLS1 as they are amongst the general Seyfert 1 population, arguing against models in which the relatively small broad-line widths of NLS1 are due to a preferred face-on orientation.

Prompted by the clear prediction of our model that equatorially- and polar-scatttered objects should have polarization position angles parallel and perpendicular, respectively, to the projected torus axis, we have re-examined the question of the relative orientations of the polarization **E** vector and the radio source axis in Seyfert 1 galaxies. In cases where the radio axis can be determined, the position angles are consistent with our two-component scattering model in that the polarization PA is generally perpendicular to the radio axis for objects exhibiting signatures of polar scattering but parallel to it in those with equatorial scattering characteristics. In common with previous studies, we find a preference for the polarization PA in Seyfert 1 galaxies to be roughly aligned with the radio source axis (within 30°). In the samples we have considered, parallel orientations outnumber perpendicular ones *($\Delta PA > 60°$)* by roughly 2:1. However, the distribution of *$\Delta PA$* is not bi-modal, intermediate values are also common. In at least one object exhibiting spectropolarimetric signatures of polar scattering, NGC 3227, the misalignment between the optical polarization PA and the perpendicular to the radio axis appears to be the result of a partially-filled scattering cone. Whether such misalignments are due to this, or other causes, the perception (e.g. Antonucci 2001) that most Seyfert 1 galaxies exhibit optical polarization parallel to the radio axis, with a few exceptions showing perpendicular orientations, appears to be an over-simplification.

We have shown that, with the addition of a compact, disc-like equatorial scattering region located within the torus, the range of optical polarization properties (particularly variations in $p$ and $\theta$ associated with the broad H$\alpha$ line) exhibited by Seyfert nuclei can be broadly explained within the context of the geometry postulated by the unification scheme. However, we have also demonstrated in this paper that the polarization in a significant fraction of Seyfert 1 galaxies is dominated by polar scattering, as in Seyfert 2 galaxies in which broad lines are detected in polarized flux. We have argued that these 'polar-scattered' Seyfert 1 galaxies are viewed through the relatively thin atmosphere of the circum-nuclear torus and are thus subject to a moderate amount of extinction. It follows that, in detail, the Seyfert phenomenon cannot accurately be described in terms of a simple '2-bin' (i.e. Seyfert 1 or Seyfert 2) orientation scheme. We require a more sophisticated model in which observed characteristics vary progressively with inclination.

## ACKNOWLEDGEMENTS

JES acknowledges financial support from PPARC. The WHT is operated on the island of La Palma by the Isaac Newton Group in the Spanish Observatorio del Roque de los Muchachos of the Instituto de Astrofisica de Canarias. The work reported in this paper was partly carried out using facilities and software provided by the Starlink project. This research has made use of the NASA/IPAC Extragalactic Database (NED), which is operated by the Jet Propulsion Laboratory, California Institute of Technology, under contract with the National Aeronautics and Space Administration.